\documentclass[11pt,draftclsnofoot,onecolumn]{IEEEtran}

\usepackage{amsmath}
\usepackage{amsfonts}
\usepackage{amssymb}
\usepackage{graphicx}
\usepackage{dsfont}
\usepackage{subfigure}
\usepackage{setspace}
\usepackage{cite}

\newcommand{\tr}{\mathop{\mathrm{tr}}\nolimits}

\newtheorem{Remark}{Remark}

\def\cA{\mathcal{A}}
\def\cB{\mathcal{B}}

\def\cF{\mathcal{F}}
\def\cG{\mathcal{G}}

\def\cK{\mathcal{K}}

\def\cM{\mathcal{M}}

\def\cP{\mathcal{P}}

\def\cS{\mathcal{S}}

\def\bcH{\boldsymbol{\mathcal{H}}}

\def\bh{{\mathbf{h}}}

\def\bn{{\mathbf{n}}}

\def\bx{{\mathbf{x}}}
\def\by{{\mathbf{y}}}

\def\b0{{\mathbf{0}}}

\def\bA{{\mathbf{A}}}
\def\bB{{\mathbf{B}}}

\def\bE{{\mathbf{E}}}
\def\bF{{\mathbf{F}}}
\def\bG{{\mathbf{G}}}
\def\bH{{\mathbf{H}}}
\def\bI{{\mathbf{I}}}

\def\bR{{\mathbf{R}}}

\def\bU{{\mathbf{U}}}
\def\bV{{\mathbf{V}}}
\def\bW{{\mathbf{W}}}

\def\bbC{{\mathbb{C}}}

\def\bbE{{\mathbb{E}}}
\def\bbF{{\mathbb{F}}}

\def\bbR{{\mathbb{R}}}

\def\bbV{{\mathbb{V}}}
\def\bbW{{\mathbb{W}}}

\def\rA{\mathrm{A}}

\def\rF{\mathrm{F}}

\def\rR{\mathrm{R}}

\def\rT{\mathrm{T}}

\def\rX{\mathrm{X}}

\def\rm{\mathrm{m}}

\def\rMMSE{\mathrm{MMSE}}

\def\rsum{\mathrm{sum}}

\def\ropt{\mathrm{opt}}

\def\cOP{\mathcal{OP}}
\def\cSP{\mathcal{SP}}

\begin{document}

\title{Joint Transmit Precoding for the Relay Interference Broadcast Channel}
\author{
Kien~T.~Truong,~\IEEEmembership{Student Member,~IEEE,} and Robert~W.~Heath,~Jr.*,~\IEEEmembership{Fellow,~IEEE}
\thanks{The authors are with the Department of Electrical and Computer Engineering, 1 University Station, C0806, The University of Texas at Austin, Austin, TX, 78712-0240 (email: kientruong@utexas.edu and rheath@ece.utexas.edu, phone: (512) 686 8225, fax: (512) 471 6512). R. Heath is the corresponding author.}
\thanks{This work was supported by a gift from Huawei Technologies, Inc.}}
{}\maketitle

\begin{abstract}
Relays in cellular systems are interference limited. The highest end-to-end sum rates are achieved when the relays are jointly optimized with the transmit strategy. Unfortunately, interference couples the links together making joint optimization challenging. Further, the end-to-end multi-hop performance is sensitive to rate mismatch, when some links have a dominant first link while others have a dominant second link. This paper proposes an algorithm for designing the linear transmit precoders at the transmitters and relays of the relay interference broadcast channel, a generic model for relay-based cellular systems, to maximize the end-to-end sum-rates. First, the relays are designed to maximize the second-hop sum-rates. Next, approximate end-to-end rates that depend on the time-sharing fraction and the second-hop rates are used to formulate a sum-utility maximization problem for designing the transmitters. This problem is solved by iteratively minimizing the weighted sum of mean square errors. Finally, the norms of the transmit precoders at the transmitters are adjusted to eliminate  rate mismatch. The proposed algorithm allows for distributed implementation and has fast convergence. Numerical results show that the proposed algorithm outperforms a reasonable application of single-hop interference management strategies separately on two hops.
\end{abstract}


\section{Introduction}\label{sec:Intro}
Multi-hop communication is one strategy to improve reliability on wireless communication links. The idea is to send information through one or more relays, who receive and retransmit from source to destination. In cellular system designs that support relaying, the relay is usually a piece of fixed infrastructure connected to a power source but not a wired backhaul. Unfortunately, aside from simple repeaters, relays have yet to see wide deployment commercial despite a tremendous amount of research~\cite{Meulen1971:AAP,Drucker1988:VTC,PabstEtAl2004:CM,BergerEtAl2009:CM,SendonarisEtAl2003a:TIT03,SendonarisEtAl2003b:TIT03,LanemanEtAl2004:TIT04,LeHossain2007:CM,SydirTaori2009:CM,YangEtAl2009:CM,3GPPLTEAdvanced:STANDARD,IEEE80216m:STANDARD}. A main reason is that relays in cellular systems are sensitive to interference~\cite{PetersEtAl2009:JWCN,WirthEtAl2009:VTC,ViswanathanMukherjee2005:TWC,YilmazEtAl2009:SPAWC}.

In this paper, we propose an algorithm for relaying in multiple-antenna relay interference channels. We assume half-duplex decode-and-forward (DF) relays~\cite{CoverGamal1979:TIT} that cannot transmit and receive at the same time. Thus, the transmission procedure consists of two stages. In the first stage, the transmitters send data to the relays. By assumption, the relays attempt to decode only the signals intended for their associated receivers. In the second stage, the relays re-generate and re-encode signals intended to each of their associated receivers before retransmitting to the receivers. While each receiver intends to receive data from only a single transmitter via a single relay, each transmitter can have independent data for multiple receivers. Consequently, each transmitter may ask for the aid of multiple relays at the same time and each relay may simultaneously forward independent data from a transmitter to multiple receivers.

Using shared radio resources, the transmissions from the transmitters to relays interfere with each other. Similarly, the transmissions from the relays to receiver interfere with each other. If considered separately, each stage of transmission is an instance of the single-hop interference broadcast channel~\cite{ShiEtAl2011:TSP,RazaviyaynEtAl2011:CISS}, a generalization of the conventional single-hop interference channel. Each transmitter in the broadcast channel has independent data for multiple receivers while each transmitter in the conventional channel has data for only one receiver. Although recent results on the single-hop interference broadcast channel~\cite{ShiEtAl2011:TSP,RazaviyaynEtAl2011:CISS} can be applied separately for the two hops, in this paper we show that higher end-to-end sum rates can be achieved when the transmitters and relays are configured jointly. Unfortunately, jointly configuring the transmitters and relays is challenging, especially with limited information about the interferers.

In general, the optimal transmit and receive strategies for sum-rate maximization in interference channels are not widely known, even for single-hop channels. Thus, we adopt a pragmatic approach that treats interference as noise and maximizes end-to-end sum-rates by searching within the class of linear transmit and receive strategies. Assuming the receivers always use the optimal linear MMSE receive filters, we focus on designing the transmit precoders at the transmitters and relays. Unfortunately, the problem is nonconvex and NP-hard. Moreover, by definition, the end-to-end achievable rates of two-hop links are not continuously differentiable at every point. Thus, finding the stationary points of the problem, including its globally and locally optimal solutions, is challenging~\cite{KoLin1995:BOOK,LiuEtAl2010:ICASSP,JorswieckEtAl2008:TSP}.

We assume all two-hop links have a common timesharing value, i.e., the same fractions of time for transmission on two hops because decode-and-forward is assumed. The achievable end-to-end rate corresponding to a relay is defined as the minimum of the achievable normalized rate from its associated transmitter to itself and the achievable normalized sum-rates from itself to its associated receivers~\cite{ChakrabartiEtAl2005:SPAWC}. A two-hop rate mismatch occurs when some links have a dominant first hop while others have a dominant second hop, resulting in low end-to-end sum-rates. An efficient system design should not cause any two-hop rate mismatch while mitigating interference.

Transmit precoder design has been studied widely for the multiple-antenna single-hop interference broadcast channel~\cite{ShiEtAl2011:TSP,RazaviyaynEtAl2011:CISS}, especially for its special case of the single-hop interference channel~\cite{ShiEtAl2011:TSP,RazaviyaynEtAl2011:CISS,SchmidtEtAl2009:Asilomar,HuangEtAl2006:JSAC,ShiEtAl2008:Allerton,ShiEtAl2009:ICC,ShiEtAl2009:MILCOM}. The methods in~\cite{HuangEtAl2006:JSAC,ShiEtAl2008:Allerton,ShiEtAl2009:ICC,ShiEtAl2009:MILCOM} were based on the so-called interference pricing framework where the transmitters configure themselves based on interference prices fed back from the receivers. Interference prices represent the marginal decrease in the sum-utility function per unit increase in interference power.  In~\cite{SchmidtEtAl2009:Asilomar,ShiEtAl2011:TSP,RazaviyaynEtAl2011:CISS}, based on a relationship between mutual information and mean squared error (MSE), sum-utility maximization problems were solved via iterative minimization of weighted sum-MSEs. Under certain conditions on the utility functions, the algorithms in \cite{HuangEtAl2006:JSAC,ShiEtAl2008:Allerton,ShiEtAl2009:ICC,ShiEtAl2009:MILCOM,SchmidtEtAl2009:Asilomar,ShiEtAl2011:TSP,RazaviyaynEtAl2011:CISS} are guaranteed to converge to the stationary points of the corresponding sum-utility maximization problems. Note that the existing single-hop results are designed specifically for the single-hop interference channel. It is not straightforward to incorporate the special features of the relay interference channel, e.g., relay signal processing operation and multi-hop transmission.

To the best of our knowledge, there has been little prior work on transmit precoder design in the relay interference broadcast channel. Prior work focused on interference mitigation for special cases of this DF relay model~\cite{ChaeEtAl2008:TSP,DartmannEtAl2011:GLOBECOM,ZhangLetaief2012:TWC,TruongHeath2011:Globecom}. In~\cite{ChaeEtAl2008:TSP}, the authors proposed a transmit precoder design for the MIMO relay broadcast channel where a single MIMO DF relay forwards data from a single transmitter to multiple receivers. This means that there is no interference on the first hop and there is no inter-relay interference on the second hop. Prior work in~\cite{DartmannEtAl2011:GLOBECOM,ZhangLetaief2012:TWC,TruongHeath2011:Globecom} considered the DF relay interference channel where the receivers are equipped with a single antenna and each relay is dedicated to aiding a single transmitter-receiver pair. The transmit precoders at the transmitters and relays are designed completely independently in~\cite{DartmannEtAl2011:GLOBECOM} to maximize the minimum average signal-to-interference-plus-noise ratio (SINR) on each hop. Prior work in~\cite{ZhangLetaief2012:TWC} investigated the end-to-end sum-rate performance of different relay cooperation strategies and found that the zero-forcing based approach is attractive for interference management. Based on the interference pricing framework, the algorithm in our prior work in~\cite{TruongHeath2011:Globecom} used approximations of end-to-end rates to compute interference prices for designing the second-hop transmit precoders with fixed first-hop transmit precoders. It is not straightforward to extend the results of~\cite{TruongHeath2011:Globecom} to the general relay interference broadcast channel with multiple-antenna receivers. There have been other algorithms for designing transmit precoders at the relays and/or transmitters in the relay interference channel~\cite{AbeEtAl2006:EURASIP,RankovWittenben2007:JSAC,DehkordyEtAl2009:TSP,ChenEtAl2009:ICASSP,NguyenEtAl2009:Globecom,LiuPetropulu2010:ICASSP,ZhangEtAl2009:TSP,NouraniEtAl2009:ISIT,GomadamJafar2010:TC,NouraniEtAl2010:ISIT,ChaliseVandendorpe2010:TSP,NingEtAl2010:ISIT,PetersEtAl2009:JWCN,PanahEtAl2011:JASP,JoungEtAl2010:TSP}. Nevertheless, much of the prior work considered either AF relays~\cite{AbeEtAl2006:EURASIP,RankovWittenben2007:JSAC,DehkordyEtAl2009:TSP,ChenEtAl2009:ICASSP,NguyenEtAl2009:Globecom,LiuPetropulu2010:ICASSP,ZhangEtAl2009:TSP,NouraniEtAl2009:ISIT,GomadamJafar2010:TC,NouraniEtAl2010:ISIT,ChaliseVandendorpe2010:TSP,NingEtAl2010:ISIT} or other relay architectures, like the shared relay~\cite{PetersEtAl2009:JWCN,PanahEtAl2011:JASP} or two-way relay~\cite{JoungEtAl2010:TSP}. In addition, in this prior work, each relay simultaneously forwards data for multiple transmitter-receiver pairs unlike in our approach.

In this paper, we propose a cooperative algorithm for efficiently finding suboptimal solutions of the transmit precoder design problem with high end-to-end sum-rates. The proposed algorithm can be implemented in a distributed fashion with low communication overhead. The algorithm consists of three phases in the following order: i) second-hop transmit precoder design, ii) first-hop transmit precoder design, and iii) first-hop transmit power control. In the first phase, we ignore the first hop and focus on configuring the relays to maximize the achievable second-hop sum-rates. Essentially, the second hop is treated as the conventional single-hop interference broadcast channel. Thus, existing single-hop algorithms can be applied to find the stationary points of second-hop sum-rate maximization~\cite{ShiEtAl2011:TSP,RazaviyaynEtAl2011:CISS,JorswieckEtAl2008:TSP,LarssonJorswieck2008:JSAC,ShiEtAl2008:Allerton,FarrokhiEtAl1998:JSAC,HuangEtAl2006:JSAC,ShiEtAl2009:ICC,ShiEtAl2009:MILCOM,ZakhourEtAl2009:VTC,HoGesbert2010:ICC,SchmidtEtAl2009:Asilomar,GomadamEtAl2011:TIT,PetersHeath2011:TVT}. Having computed the second-hop transmit precoders, each relay computes the sum of achievable rates from itself to its associated receivers, which is used as input to the second phase.

The second phase focuses on designing the first-hop transmit precoders. In the naive approach, this can be done by applying the prior work in~\cite{ShiEtAl2011:TSP,RazaviyaynEtAl2011:CISS,JorswieckEtAl2008:TSP,LarssonJorswieck2008:JSAC,ShiEtAl2008:Allerton,FarrokhiEtAl1998:JSAC,HuangEtAl2006:JSAC,ShiEtAl2009:ICC,ShiEtAl2009:MILCOM,ZakhourEtAl2009:VTC,HoGesbert2010:ICC,SchmidtEtAl2009:Asilomar,GomadamEtAl2011:TIT,PetersHeath2011:TVT} while ignoring the designed second-hop transmit precoders. The naive approach, however, may cause a two-hop rate mismatch because it does not take into account the timesharing value and second-hop configuration. To overcome this limitation, we formulate and solve a new problem to maximize an approximation of the achievable end-to-end sum-rates. Such approximations of the achievable end-to-end rates depend not only on the first-hop transmit precoders, but also on the timesharing value and second-hop configuration. This allows for second-hop interference mitigation at the same time as rate-mismatch alleviation. Some guidelines for selecting such approximations are provided. Having defined a more comprehensive utility function, we use the technique in~\cite{SchmidtEtAl2009:Asilomar,ShiEtAl2011:TSP,RazaviyaynEtAl2011:CISS} to develop an iterative method that is guaranteed to converge to the stationary points of the new sum-utility maximization problem. This concludes the second phase of the proposed algorithm.

The output of the second phase may contain a residual two-hop rate mismatch since only an approximate solution is proposed. In the final phase, we propose to fix the shapes of the first-hop transmit precoders and to adjust their norms to eliminate completely two-hop rate mismatching. Essentially, this is a transmit power control problem. Note that for a two-hop link with a dominant first hop, excess power is allocated for the transmissions on the first hop. We propose a method for simultaneously reducing excess power for the first-hop transmissions so that the achievable end-to-end rates for all the relays are nondecreasing over iterations, thus potentially improving the achievable end-to-end sum-rates. The method is guaranteed to converge. At the convergence point, there are no two-hop links with a dominant first hop, i.e., there is no longer two-hop rate mismatch. Although there have been many power control algorithms for the single-hop interference channel~\cite{HuangEtAl2006:JSAC,ShiEtAl2009:ISIT,FoschiniMiljanic1993:TVT,Yates1995:JSAC,SaraydarEtAl2002:TC,SungWong2003:TWC,StanczakEtAl2007:TSP,SuVanDerSchaar2009:TWC,QianEtAl2009:TWC,StanczakEtAl2009:BOOK} and for the relay interference channel~\cite{ShiEtAl2009:TWC,ZhouEtAl2009:ICC,ShiEtAl2010:ICC,XiaoCuthbert2009:VTC}, they are not designed to eliminate two-hop rate mismatching. Therefore, even if applicable to the third phase, existing power control algorithms may worsen the two-hop rate mismatch situation.

We use Monte Carlo simulations to evaluate the average achievable end-to-end sum-rates of the proposed algorithm for various relay interference broadcast channel configurations. The na\"{i}ve approach of applying existing single-hop results separately for the two hops is selected as the baseline strategy. The proposed algorithm and the na\"{i}ve approach have the same second-hop transmit precoders. While the timesharing value and second-hop configuration are taken into account in the first-hop transmit precoder design in the last two phases of the proposed algorithm, they are ignored in that of the na\"{i}ve approach. Numerical results show that the first two phases of the proposed algorithm are enough to provide large end-to-end sum-rate gains over the na\"{i}ve approach. In addition, the last phase of the proposed algorithm makes considerable improvements in end-to-end sum-rate performance over the output of the second phase. This highlights the importance of two-hop rate matching in the DF relay interference (broadcast) channel. Finally, each phase of the proposed algorithm is observed to converge in a few iterations. Note that the proposed algorithm can be implemented in a  distributed manner with a little more overhead than the na\"{i}ve approach. Thus, it is suitable for practical implementation in DF relay networks.

The organization of the remainder of this paper is as follows. Section~\ref{sec:systemModel} describes the system model. Section~\ref{sec:problemFormulation} formulates the design problem and discusses the challenges.  Section \ref{sec:proposedApproach} presents the proposed algorithm in detail. Section~\ref{sec:simulation} numerically evaluates the achievable end-to-end sum-rates of the proposed algorithm. Section~\ref{sec:conclusion} concludes this paper and suggests future research.

Notation: We use normal letters (e.g., $a$) for scalars, lowercase and uppercase boldface letters (e.g., $\bh$ and $\bH$) for column vectors and matrices. $\bI_{N}$ is the identity matrix of size $N \times N$. For a matrix $\bA$, $\bA^{*}$ is the conjugate transpose, $\tr(\bA)$ the trace, $|\bA|$  the determinant, and $\|\bA\|_{\rF}$ the Frobenius norm. $\bbE[\cdot]$ is the statistical expectation operator. $()^{(n)}$ denotes iteration index. $()_{\rT}$ is used for transmitters' parameters, $()_{\rR}$ for receivers', and $()_{\rX}$ for relays'.

\section{System Model}\label{sec:systemModel}
\begin{figure}[h]
\centering
\includegraphics[width=3.4in]{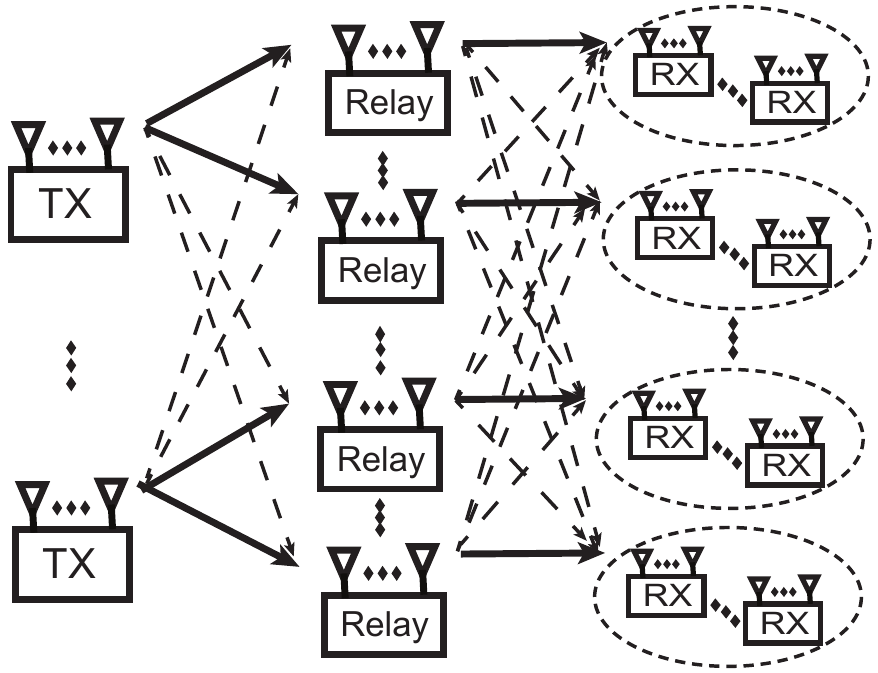}
\caption{A relay interference broadcast channel where a number of half-duplex DF relays aid the data transmission from a number of transmitters (TXs) to their associated receivers (RXs). The solid lines connect the associated nodes and represent communication links. The dashed lines represent interference links.}
\label{fig:MIMORelayIntfBroadcastChannel}
\end{figure}

Consider a relay interference broadcast channel where $K_{\rT}$ transmitters communicate with $K_{\rR}$ receivers with the aid of $K_{\rX}$ half-duplex DF relays, as illustrated in Fig. \ref{fig:MIMORelayIntfBroadcastChannel}. Each transmitter is assigned a unique index from $\cK_{\rT} \triangleq \{1, \cdots, K_{\rT}\}$. Similarly, each relay is assigned a unique index from $\cK_{\rX} \triangleq \{1, \cdots, K_{\rX}\}$ and each receiver is assigned a unique index from $\cK_{\rR}\triangleq \{1, \cdots, K_{\rR}\}$. Each transmitter may require the aid of multiple relays to simultaneously send independent data streams to its receivers. Each relay is dedicated to serving multiple receivers that are associated with a single transmitter. Each receiver intends to receive data from only one transmitter with the aid of a single relay. Let $\chi(k) \in \cK_{\rX}$ denote the index of the relay that aids receiver $k \in \cK_{\rR}$. Let $\mu(k) \in \cK_{\rT}$ denote the index of the transmitter that is aided by relay $k \in \cK_{\rX}$. The transmitters and relays do not share data. We assume that each relay $k$ does not attempt to decode the signal intended for receiver $m \in \cK_{\rR}$ with $\chi(m) \neq k$. Transmitter $k \in \cK_{\rT}$ has $N_{\rT,k}$ antennas, relay $m \in \cK_{\rX}$ has $N_{\rX,m}$ antennas, and receiver $q \in \cK_{\rR}$ has $N_{\rR,q}$ antennas. 

Half-duplex relays cannot transmit and receive at the same time, thus the transmission procedure requires two stages. Using a common frequency at the same time, the transmissions in the same stage interfere with each other. For tractable analysis, we assume Gaussian signaling is used in both stages although it may not be optimal for the relay interference broadcast channel. In the first stage, the transmitters send data to the relays. Treating unwanted signals as additive Gaussian noise, each relay decodes its desired signal. Each relay separates its decoded signals for individual associated receivers, re-encodes and retransmits to its associated receivers in the second stage. Each receiver treats unwanted signals as additive Gaussian noise when decoding its desired signal.

We consider slowly-varying, frequency-flat, block-fading channels. We denote $\bH_{m,k} \in \bbC^{N_{\rX,m} \times N_{\rT,k}}$ as the matrix channel between transmitter $k$ and relay $m$ for $k \in \cK_{\rT}$ and $m \in \cK_{\rX}$. Let $\bx_{1,k} \in \bbC^{d_{1, k} \times 1}$ denote the symbol vector that transmitter $\mu(k) \in \cK_{\rT}$ sends to relay $k \in \cK_{\rX}$ in the first stage, where $d_{1,k}$ is the number of data streams and $\bbE(\bx_{1,k}\bx^{*}_{1,k}) = \bI_{d_{1, k}}$. We denote $\bF_{\rT,k} \in \bbC^{N_{\rT,\mu(k)} \times d_{1,k}}$ as the linear transmit precoder that transmitter $\mu(k)$ uses to send $\bx_{1,k}$ to relay $k \in \cK_{\rX}$. We define $\bF_{\rT} \triangleq \{\bF_{\rT, 1}, \cdots, \bF_{\rT, K_{\rX}}\} \in \bbF_{\rT} \triangleq \bbC^{N_{\rT,\mu(1)} \times d_{1,1}} \times \cdots \times \bbC^{N_{\rT,\mu(K_{\rX})} \times d_{1,K_{\rX}}}$. Let $p_{\rT,k}$ be the sum transmit power at transmitter $k \in \cK_{\rT}$. The sum transmit power constraint at transmitter $k \in \cK_{\rT}$ is
\begin{equation}\label{eq:PowerConstraint1}
\sum_{\substack{m \in \cK_{\rX}\\\mu(m) = k}}\|\bF_{\rT,m}\|^{2}_{\rF} = \sum_{\substack{m \in \cK_{\rX}\\\mu(m) = k}} \tr(\bF_{\rT,m}\bF^{*}_{\rT,m}) \leq p_{\rT,k}, \forall k \in \cK_{\rT}.
\end{equation}
We denote $\bn_{\rX,k} \in \bbC^{N_{\rX,k} \times 1}$ as spatially white, additive Gaussian noise at relay $k \in \cK_{\rX}$ with $\bbE(\bn_{\rX,k}\bn^{*}_{\rX,k}) = \sigma^{2}_{\rX,k}\bI_{N_{\rX,k}}$. Relay $k \in \cK_{\rX}$ observes
\begin{equation}
\by_{\rX,k} = \underbrace{\bH_{k,\mu(k)}\bF_{\rT,k}\bx_{1,k}}_{\mathrm{desired~signal}} +
\underbrace{\sum_{\substack{m \in \cK_{\rX}\\ m\neq k}}\bH_{k,\mu(m)}\bF_{\rT,m}\bx_{1,m}}_{\mathrm{interference}} + \bn_{\rX,k}.\label{eq:y1}
\end{equation}
The interference plus noise covariance matrix at relay $k \in \cK_{\rX}$ is
\begin{equation}
\bR_{\rX,k}(\bF_{\rT}) =  \sum_{\substack{m \in \cK_{\rX}\\m\neq k}} \bH_{k,\mu(m)}\bF_{\rT,m}\bF^{*}_{\rT,m}\bH^{*}_{k,\mu(m)} + \sigma^{2}_{\rX,k}\bI_{N_{\rX,k}}.
\end{equation}
Each relay $k \in \cK_{\rX}$ applies a linear receive filter $\bW_{\rX,k} \in \bbC^{N_{\rX,k} \times d_{1,k}}$ to $\by_{\rX,k}$. We define $\bW_{\rX} \triangleq (\bW_{\rX,1}, \cdots, \bW_{\rX,K_{\rX}}) \in \bbW_{\rX} \triangleq \bbC^{N_{\rX,1} \times d_{1,1}} \times \cdots \times \bbC^{N_{\rX,K_{\rX}} \times d_{1,K_{\rX}}}$. The maximum achievable rate on the first hop from transmitter $\mu(k) \in \cK_{\rT}$ to relay $k \in \cK_{\rX}$ is
\begin{eqnarray}
R_{1,k}(\bF_{\rT}) &=&  \log_{2}\det\left(\bI+ \bF^{*}_{\rT,k}\bH^{*}_{k,\mu(k)}\left[\bR_{\rX,k}(\bF_{\rT})\right]^{-1}\bH_{k,\mu(k)}\bF_{\rT,k}\right),
\end{eqnarray}
which can be achieved using the linear MMSE receive filter
\begin{equation}\label{eq:MMSEreceiveFilter}
\bW^{\rMMSE}_{\rX,k} = \left[\bH_{k,\mu(k)}\bF_{\rT,k}\bF^{*}_{\rT,k}\bH^{*}_{k,\mu(k)} + \bR_{\rX,k}(\bF_{\rT})\right]^{-1}\bH_{k,\mu(k)}\bF_{\rT,k}.
\end{equation}
We define $\xi_{1,k}(\bF_{\rT}) \triangleq 2^{R_{1,k}(\bF_{\rT})} - 1$. Intuitively, $\xi_{1,k}(\bF_{\rT})$ can be thought of as the effective first-hop SINR to relay $k$ if we send a single data stream from transmitter $\mu(k) \in \cK_{\rT}$ to relay $k$ at the rate $R_{1,k}(\bF_{\rT})$.

Let $\bG_{m,k} \in \bbC^{N_{\rR,m} \times N_{\rX,k}}$ denote the matrix channel between relay $k \in \cK_{\rX}$ and receiver $m \in \cK_{\rR}$. Let $\bx_{2,m} \in \bbC^{N_{\rX,\chi(m)} \times d_{2,m}}$ denote the transmit symbol vector that relay $\chi(m) \in \cK_{\rX}$ sends to receiver $m \in \cK_{\rR}$ where $d_{2,m}$ is the number of data streams and $\bbE(\bx_{2,m}\bx^{*}_{2,m}) = \bI_{d_{2,m}}$. We denote $\bF_{\rX,m} \in \bbC^{N_{\rX,\chi(m)} \times d_{2,m}}$ as the linear transmit precoder that relay $\chi(m)$ uses to send $\bx_{2,m}$ to receiver $m \in \cK_{\rR}$. We define $\bF_{\rX} \triangleq \{\bF_{\rX, 1}, \cdots, \bF_{\rX, K_{\rR}}\} \in \bbF_{\rX} \triangleq \bbC^{N_{\rX,\chi(1)} \times d_{2,1}} \times \cdots \times \bbC^{N_{\rX,\chi(K_{\rR})} \times d_{1,K_{\rR}}}$. Let $p_{\rX,k}$ be the sum transmit power at relay $k \in \cK_{\rX}$. The sum transmit power constraint at relay $k \in \cK_{\rX}$ is
\begin{equation}\label{eq:PowerConstraint2}
\sum_{\substack{m \in\cK_{\rR}\\\chi(m) = k}} \|\bF_{\rX,m}\|_{\rF}^{2} = \sum_{\substack{m \in\cK_{\rR}\\\chi(m) = k}} \tr(\bF_{\rX,m}\bF^{*}_{\rX,m}) \leq p_{\rX,k}, \forall k \in \cK_{\rX}.
\end{equation}
 We denote $\bn_{\rR,m} \in \bbC^{N_{\rR,m}\times 1}$ as spatially white, additive Gaussian noise at receiver $m \in \cK_{\rR}$ with $\bbE(\bn_{\rR,m}\bn^{*}_{\rR,m}) = \sigma^{2}_{\rR,m}\bI_{N_{\rR,m}}$. Receiver $m \in \cK_{\rR}$ observes
\begin{equation}
\by_{\rR,m} = \underbrace{\bG_{m,\chi(m)}\bF_{\rX,m}\bx_{2,m}}_{\mathrm{desired~signal}} +\underbrace{\sum_{\substack{q \in \cK_{\rR}\\q \neq m}} \bG_{m,\chi(q)}\bF_{\rX,q}\bx_{2,q}}_{\mathrm{interference}} + \bn_{\rR,m}.\label{eq:y2}
\end{equation}
The interference plus noise covariance matrix at receiver $m \in \cK_{\rR}$ is
\begin{equation}
\bR_{\rR,m}(\bF_{\rX}) =  \sum_{\substack{q \in \cK_{\rR}\\q \neq m}} \bG_{m,\chi(q)}\bF_{\rX,q}\bF^{*}_{\rX,q}\bG^{*}_{m,\chi(q)} + \sigma^{2}_{\rR,m}\bI_{N_{\rR,m}}.
\end{equation}
Each receiver $m \in \cK_{\rR}$ applies a linear receive filter $\bW_{\rR,m} \in \bbC^{N_{\rR,m} \times d_{2,m}}$ to $\by_{\rR,m}$. The maximum achievable rate at receiver $m \in\cK_{\rR}$ is
\begin{equation}
R_{2,m}(\bF_{\rX}) =  \log_{2}\det\left(\bI + \bF^{*}_{\rX,m}\bG^{*}_{m,\chi(m)}\left[\bR_{\rR,m}(\bF_{\rX})\right]^{-1}\bG_{m,\chi(m)}\bF_{\rX,m}\right),
\end{equation}
which can be achieved by the following linear MMSE receive filter
\begin{equation}
\bW^{\rMMSE}_{\rR,m} = \left[\bG_{m,\chi(m)}\bF_{\rX,m}\bF^{*}_{\rX,m}\bG^{*}_{m,\chi(m)} + \bR_{\rR,m}(\bF_{\rX})\right]^{-1}\bG_{m,\chi(m)}\bF_{\rX,m}.
\end{equation}
The sum of maximum achievable second-hop rates for relay $k \in \cK_{\rX}$ is defined as
\begin{eqnarray}
R_{2,k, \rsum}(\bF_{\rX}) &\triangleq & \sum_{\substack{m\in \cK_{\rR}\\\chi(m) = k}} R_{2,m}(\bF_{\rX}).\label{eq:RsumDef}
\end{eqnarray}
We define $\xi_{2,k}(\bF_{\rX}) \triangleq 2^{R_{2,k, \rsum}(\bF_{\rX})}-1$. Similar to $\xi_{1,k}(\bF_{\rT})$, we refer to $\xi_{2,k}(\bF_{\rX})$ as the effective SINR on the second hop corresponding to relay $k$.

We assume the transmissions in each stage start and end at the same time. Let $t$ be the fraction of time for transmission on the first hop, which is also referred to as the timesharing value. The fraction of time for transmission on the second hop is $(1-t)$. For example, in 3GPP LTE-Advanced cellular systems, $t$ depends on the number of subframes allocated to the backhaul links (i.e., between base stations and relays) in a radio frame~\cite{3GPPLTEAdvanced:STANDARD}. In this paper, $t$ is a fixed parameter. The optimization of the timesharing value is left for future work. For relay $k \in \cK_{\rX}$, the normalized  rate on the first hop is $tR_{1,k}(\bF_{\rT})$ while the normalized sum of second-hop rates for this relay is $(1-t)R_{2,k,\rsum}(\bF_{\rX})$. Based on the relative comparison of the normalized rates on two hops, the two-hop link corresponding to relay $k \in \cK_{\rX}$ can be classified into the following three categories: i) first-hop dominant if $tR_{1,k}(\bF_{\rT}) > (1-t)R_{2,k,\rsum}(\bF_{\rX})$, ii) second-hop dominant if $tR_{1,k}(\bF_{\rT}) < (1-t)R_{2,k,\rsum}(\bF_{\rX})$, and iii) equal rate if $tR_{1,k}(\bF_{\rT}) = (1-t)R_{2,k,\rsum}(\bF_{\rX})$.

The achievable end-to-end rate for relay $k \in \cK_{\rX}$ is defined as the minimum of the normalized achievable rates on two hops~\cite{ChakrabartiEtAl2005:SPAWC}
\begin{eqnarray}
R_{k}(\bF_{\rT},\bF_{\rX}) &\triangleq&\min\{tR_{1,k}(\bF_{\rT}), (1-t)R_{2,k,\rsum}(\bF_{\rX})\}\label{relayIBC:eq:e2eRdef0}\\
&=&\min\{t\log_{2}(1 + \xi_{1,k}(\bF_{\rT})), (1 - t)\log_{2}(1 + \xi_{2,k}(\bF_{\rR}))\}.\label{relayIBC:eq:e2eRdef}
\end{eqnarray}
Given the achievable rates on two hops, the end-to-end achievable rate for relay $k$ can be obtained by properly allocating fractions of the normalized first-hop rate $tR_{1,k}(\bF_{\rT})$ for sending data for its associated receivers. Specifically, let $\beta_{k,m} \geq 0$ be the first-hop rate corresponding to receiver $m \in \cK_{\rR}$ such that $\chi(m) = k$, where
\begin{eqnarray}
\sum_{\substack{m \in \cK_{\rR}\\ \chi(m) = k}} \beta_{k,m} \leq tR_{1,k}(\bF_{\rT}).\label{eq:e2eCondition}
\end{eqnarray}
For example, we can achieve $R_{k}(\bF_{\rT},\bF_{\rX})$ in (\ref{relayIBC:eq:e2eRdef0}) if relay $k$ uses the following first-hop rate allocation strategy
\begin{eqnarray}\label{eq:beta}
\beta_{k,m} = \begin{cases}
(1-t)R_{2,m}(\bF_{\rX}), & \mbox{if~} tR_{1,k}(\bF_{\rT}) \geq (1-t)R_{2,k,\rsum}(\bF_{\rX})\\
tR_{1,k}(\bF_{\rT})\frac{R_{2,m}(\bF_{\rX})}{R_{2,k, \rsum}(\bF_{\rX})}, & \mbox{otherwise}.
\end{cases}
\end{eqnarray}
We can check that $\beta_{k,m}$ for $k \in \cK_{\rX}$ and $m \in \cK_{\rR}$ satisfies the condition in (\ref{eq:e2eCondition}). The end-to-end sum-rate is defined as
\begin{equation}
R_{\rsum}(\bF_{\rT},\bF_{\rX}) \triangleq \sum_{k \in \cK_{\rX}} R_{k}(\bF_{\rT},\bF_{\rX}).
\end{equation}
The design of $\bF_{\rT}$ and $\bF_{\rX}$ for maximizing $R_{\rsum}(\bF_{\rT},\bF_{\rX})$ should take into account $t$.

\section{Problem Formulation}\label{sec:problemFormulation}
This section formulates the design problem and discusses the challenges in finding optimal solutions. The problem of designing the transmit precoders at the transmitters and relays to maximize the sum of achievable end-to-end rates is formulated as follows
\begin{eqnarray}
(\cOP):~\max_{(\bF_{\rT},\bF_{\rX}) \in \bbF_{\rT} \times \bbF_{\rX}} && R_{\rsum}(\bF_{\rT}, \bF_{\rX})\nonumber\\
\mbox{subject~to} &&\sum_{\substack{m\in \cK_{\rR}\\\chi(m) = k}} \tr(\bF_{\rX,m}\bF^{*}_{\rX,m}) \leq p_{\rX,k}, \forall k \in \cK_{\rX}\\
&& \sum_{\substack{m\in \cK_{\rX}\\\mu(m) = k}} \tr(\bF_{\rT,m}\bF^{*}_{\rT,m}) \leq p_{\rT,k}, \forall k \in \cK_{\rT}.
\end{eqnarray}
The transmit precoder design problem for sum-rate maximization in the single-hop interference channel is nonconvex and NP-hard~\cite{LiuEtAl2010:ICASSP,JorswieckEtAl2008:TSP}. This means that its globally optimal solutions cannot be found efficiently in terms of computational complexity even in a centralized fashion. The more complicated problem, $(\cOP)$ is expected to be NP-hard as well. Moreover, $(\cOP)$ is a sum-utility maximization problem where the per-user utility function is not smooth at every point. Thus, it is challenging to find the stationary points of $(\cOP)$, including its optimal solutions~\cite{KoLin1995:BOOK}. 

In this paper, we focus on finding suboptimal solutions to $(\cOP)$ with high values of achievable end-to-end sum-rates. Remark \ref{rem:interference} and Remark \ref{rem:twohopRateMismatch} discusses two main challenges in solving for high-quality suboptimal solutions of $(\cOP)$. 

\begin{Remark}\label{rem:interference}
Interference mitigation is a challenge in end-to-end sum-rate maximization.  According to (\ref{eq:y1}), each relay observes undesired signals from unintended transmitters on the first hop. Similarly, according to (\ref{eq:y2}), each receiver observes undesired signals from unintended relays as well as from its associated relay (but they are intended for other receivers). Due to interference, there exists coupling among the achievable rates on the same hop. 
\end{Remark}

\begin{Remark}\label{rem:twohopRateMismatch}
Two-hop rate matching is a challenge in maximizing the end-to-end sum-rates of the DF relay interference broadcast channel. Specifically, for a given $t$, $\bF_{\rT}$, and $\bF_{\rR}$, there may exist a mismatch between the normalized achievable rates on two hops. By definition, a two-hop rate mismatch occurs when there exist $k, m \in \cK_{\rX}$ and $k \neq m$, such that $tR_{1,k}(\bF_{\rT}) > (1-t)R_{2,k,\rsum}(\bF_{\rX})$ and $tR_{1,m}(\bF_{\rT}) < (1-t) R_{2,m,\rsum}(\bF_{\rX})$. When this happens, it is always possible to improve the end-to-end sum-rate performance of the system design. For example, the norm of $\bF_{\rT,k}$ can be scaled down to obtain a new set of transmit precoders $(\bF'_{\rT}, \bF_{\rX})$ so that $tR_{1,k}(\bF'_{\rT}) = (1-t)R_{2,k,\rsum}(\bF_{\rX})$. This decreases the interference power from transmitter $k$ to all other relays on the first hop, improving the achievable rates to all other relays, especially $tR_{1,m}(\bF'_{\rT}) > tR_{1,m}(\bF_{\rT})$. This means that $R_{\rsum}(\bF'_{\rT}, \bF_{\rX}) > R_{\rsum}(\bF_{\rT}, \bF_{\rX})$. Thus, an efficient transmit precoder design in terms of end-to-end sum-rate maximization should not cause any two-hop rate mismatch.
\end{Remark}

\section{Transmit Beamforming Design}\label{sec:proposedApproach}
This section presents an algorithm for finding high-quality suboptimal solutions to $(\cOP)$. Section~\ref{subsec:BF2} discusses briefly the design of $\bF_{\rX}$. Section \ref{subsec:BF1} presents the design of $\bF_{\rT}$.

\subsection{Second-Hop Transmit Beamforming Design}\label{subsec:BF2}
We design $\bF_{\rX}$ by treating the transmission on the second hop as the single-hop interference broadcast channel. Specifically, we need to solve the following optimization problem
\begin{eqnarray}
(\cS\cP):~\max_{\bF_{\rX} \in \bbF_{\rX}} && \sum_{k\in \cK_{\rX}} R_{2,k,\rsum}(\bF_{\rX})\nonumber\\
\mbox{subject~to} && \sum_{\substack{m\in \cK_{\rR}\\\chi(m) = k}} \tr(\bF_{\rX,m}\bF^{*}_{\rX,m}) \leq p_{\rX,k}, \forall k \in \cK_{\rX}.
\end{eqnarray}
Note that $(\cS\cP)$ is nonconvex and NP-hard. Nevertheless, its stationary points can be found by existing algorithms for the single-hop interference broadcast channel. The principle of many existing algorithms is to formulate a series of related optimization problems that can be solved in polynomial time by available methods and provide multiple approximations or relaxations of the original sum-utility problem. In general, the globally optimal solutions of these related problems converge to the stationary points of the original sum-utility maximization problem. The key requirement for the applicability of existing algorithms is that the utility function of the original problem is continuously differentiable at every point. An example is the algorithm for transmit precoder design via matrix-weighted sum-MSE minimization in~\cite{ShiEtAl2011:TSP,RazaviyaynEtAl2011:CISS}. Due to space limits, the details of the algorithm are deferred to~\cite{ShiEtAl2011:TSP,RazaviyaynEtAl2011:CISS}.

Let $\bar{\bF}_{\rX}$ denote the resulting second-hop transmit precoders. We denote $\bar{\xi}_{2,k} = \xi_{2,k}(\bar{\bF}_{\rX})$ for $k \in \cK_{\rX}$. By setting $R_{2,k,\rsum}(\bar{\bF}_{\rX})$ equal to $\log_{2}(1+\bar{\xi}_{2,k})$, we obtain
\begin{equation}
\bar{\xi}_{2,k} = 2^{R_{2,k,\rsum}(\bar{\bF}_{\rX})} - 1.
\end{equation}
We assume that each receiver $m \in \cK_{\rR}$ feeds back the value of $R_{2,m}(\bar{\bF}_{\rX})$ to its associated relay, i.e., relay $\chi(m)$. Then, we assume that each relay $k \in \cK_{\rX}$ can compute $R_{2,k,\rsum}(\bar{\bF}_{\rX})$ from its corresponding $R_{2,m}(\bar{\bF}_{\rX})$ such that $k = \chi(m)$  and then $\bar{\xi}_{2,k}$. 

\subsection{First-Hop Transmit Beamforming Design}\label{subsec:BF1}
\subsubsection{Subproblem Formulation and Challenges}
This section focuses on designing $\bF_{\rT}$ given knowledge of $t$ and $\bar{\xi}_{2,k}$ for $k \in \cK_{\rX}$. It follows from $(\cOP)$ that the problem for designing $\bF_{\rT}$ is
\begin{eqnarray}
(\cF\cP):~\max_{\bF_{\rT} \in \bbF_{\rT}} && \sum_{k\in \cK_{\rX}} \min\{t\log_{2}(1 + \xi_{1,k}(\bF_{\rT})), (1-t)\log_{2}(1+\bar{\xi}_{2,k})\}\nonumber\\
\mbox{subject~to} && \sum_{\substack{m\in \cK_{\rX}\\\mu(m) = k}} \tr(\bF_{\rT,m}\bF^{*}_{\rT,m}) \leq p_{\rT,k}, \forall k \in \cK_{\rT}.
\end{eqnarray}
Note that $(\cF\cP)$ belongs to the same class of nonconvex and NP-hard sum-utility maximization problems as $(\cS\cP)$ and it is even more complicated than $(\cS\cP)$. While the objective function of $(\cS\cP)$ depends only on the corresponding transmit precoders, i.e., $\bF_{\rX}$, that of $(\cF\cP)$ depends not only on $\bF_{\rT}$ but also on $t$ and $\bar{\xi}_{k}$ for $k \in \cK_{\rX}$.

\begin{Remark}\label{rem:FPnoncontinouslyDifferentiable}
It is not possible to apply existing algorithms developed for the single-hop interference broadcast channel to find the stationary points of $(\cF\cP)$. As discussed in Section~\ref{subsec:BF2}, existing algorithms require that the utility function of sum-utility maximization problems be continuously differentiable at every point. Due to the $\min$ operation, however, the utility function of $(\cF\cP)$ is not continuously differentiable with respect to $\xi_{1,k}(\bF_{\rT})$ at the point that makes $t\log_{2}(1 + \xi_{1,k}(\bF_{\rT}))$ equal to $(1 - t)\log_{2}(1 + \bar{\xi}_{2,k})$.
\end{Remark}

\begin{Remark}\label{rem:naiveApproach}
In the na\"{i}ve approach, the timesharing and second-hop configuration are ignored, leading to an approximation of the objective function $\sum_{k \in \cK_{\rX}} \log_{2}(1 + \xi_{1,k}(\bF_{\rT}))$. The resulting  problem has the same form as $(\cS\cP)$. Thus, its stationary points can be found by existing single-hop algorithms.
\end{Remark}

For notational convenience, we define $\eta_{k}$ as the following function of $\bar{\xi}_{2,k}$ and $t$
\begin{equation}\label{eq:etak}
\eta_{k} = (1 + \bar{\xi}_{2,k})^{\frac{1 - t}{t}} - 1.
\end{equation}
Note that $t\log_{2}(1 + \eta_{k}) = (1 - t)\log_{2}(1 + \bar{\xi}_{2,k})$. This means that $\xi_{1,k}(\bF_{\rT})$ is equal to $\eta_{k}$ when the achievable first-hop rate at relay $k$ matches with the sum of achievable second-hop rates from relay $k$ to its associated receivers. Thus, $\eta_{k}$ is the rate-matching received SINR at  relay $k$.

\subsubsection{Proposed Approach}
It is challenging to find the stationary points of $(\cF\cP)$ because the utility function in the problem is not continuously differentiable at every point. In the section, we aim at finding suboptimal solutions to $(\cF\cP)$ with high end-to-end sum-rates. Instead of solving directly $(\cF\cP)$, we propose to formulate and solve a new sum-utility maximization problem, which we refer to as $(\cA\cF\cP)$. Having the same constraints as $(\cF\cP)$, $(\cA\cF\cP)$ uses an approximation of $\min\{t\log_{2}(1 + \xi_{1,k}(\bF_{\rT})), (1-t)\log_{2}(1+\bar{\xi}_{2,k})\}$ as the utility function. Note that such approximations depend on not  only $\bF_{\rT}$ but also $t$ and $\bar{\xi}_{2,k}$ for $k \in \cK_{\rX}$. Let $u_{k}(\xi_{1,k}(\bF_{\rT}), t, \bar{\xi}_{2,k})$ denote the utility function of $(\cA\cF\cP)$. In addition, we propose to solve $(\cA\cF\cP)$ via iterative minimization of weighted sum-MSEs, the well-established technique that has been used widely in prior work~\cite{ChristensenEtAl2008:TWC,SchmidtEtAl2009:Asilomar,RazaviyaynEtAl2011:CISS,ShiEtAl2011:TSP}.

Some guidelines for selecting $u_{k}(\xi_{1,k}(\bF_{\rT}), t, \bar{\xi}_{2,k})$ are provided. First, it must be twice continuously differentiable with respect to $\xi_{1,k}(\bF_{\rT})$ at every point. Second, it must satisfy the following condition 
\begin{equation}
(1+ \xi_{1,k}(\bF_{\rT}))u''_{k}(\xi_{1,k}(\bF_{\rT}), t, \bar{\xi}_{2,k}) + 2u'_{k}(\xi_{1,k}(\bF_{\rT}), t, \bar{\xi}_{2,k}) \geq 0.
\end{equation}
This condition is required to ensure that the resulting iterative algorithm for solving $(\cA\cF\cP)$ is convergent as shown in Proposition 1 in~\cite{SchmidtEtAl2009:Asilomar}. There are many approximate functions that satisfy the conditions in the guidelines. It is still unclear, however, what is the best approximation.

According to Remark \ref{rem:FPnoncontinouslyDifferentiable}, the utility function of $(\cF\cP)$ is not continuously differentiable with respect to $\xi_{1,k}(\bF_{\rT})$ at $\xi_{1,k}(\bF_{\rT}) = \eta_{k}$. We propose an approximation of the utility function of $(\cF\cP)$ as follows
\begin{eqnarray}
u_{k}(\xi_{1,k}(\bF_{\rT}), t, \bar{\xi}_{2,k}) = \begin{cases}
t \log_{2}(1 + \xi_{1,k}(\bF_{\rT})), \mbox{~if~} \xi_{1,k}(\bF_{\rT}) \leq \eta_{k},\\
t \log_{2}(1 + \eta_{k}) + \frac{t}{\log 2}\left[\exp\left(1 - \frac{1 + \eta_{k}}{1 + \xi_{1,k}(\bF_{\rT})}\right)  - 1\right], \mbox{~otherwise}.
\end{cases}\label{eq:uk}
\end{eqnarray}
Note that if $\xi_{1,k}(\bF_{\rT}) \leq \eta_{k}$, then $u_{k}(\xi_{1,k}(\bF_{\rT}), t, \bar{\xi}_{2,k})$ is exactly equal to $tR_{1,k}(\bF_{\rT})$ and hence equal to $R_{k}(\bF_{\rT},\bar{\bF}_{\rX})$. If $\xi_{1,k}(\bF_{\rT}) > \eta_{k}$, then $u_{k}(\xi_{1,k}(\bF_{\rT}), t, \bar{\xi}_{2,k})$ is larger than both $tR_{1,k}(\bF_{\rT})$ and $R_{k}(\bF_{\rT},\bar{\bF}_{\rX})$. Moreover, when $\xi_{1,k}(\bF_{\rT}) > \eta_{k}$, the gap between $u_{k}(\xi_{1,k}(\bF_{\rT}), t, \bar{\xi}_{2,k})$ and  $R_{k}(\bF_{\rT},\bar{\bF}_{\rX})$ increases with $\xi_{1,k}(\bF_{\rT})$ and is upper bounded by $t / \log 2$. Using the approximation $u_{k}(\xi_{1,k}(\bF_{\rT}), t, \bar{\xi}_{2,k})$ in~(\ref{eq:uk}) as the utility function, we formulate $(\cA\cF\cP)$ as follows
\begin{eqnarray}
(\cA\cF\cP):~\max_{\bF_{\rT} \in \bbF_{\rT}} && \sum_{k \in \cK_{\rX}} u_{k}(\xi_{1,k}(\bF_{\rT}), t, \bar{\xi}_{2,k})\nonumber\\
\mbox{subject~to} && \sum_{\substack{m\in \cK_{\rX}\\\mu(m) = k}} \tr(\bF_{\rT,m}\bF^{*}_{\rT,m}) \leq p_{\rT,k}, \forall k \in \cK_{\rT}.\end{eqnarray}
The stationary points of $(\cA\cF\cP)$ are expected to correspond to suboptimal solutions to $(\cF\cP)$.

\subsubsection{Sum-Utility Maximization via Matrix-Weighted Sum-MSE Minimization}\label{subsubsec:sumMMSE}
We develop an algorithm for solving $(\cA\cF\cP)$ via an iterative minimization of weighted sum-MSEs. We denote $\bE_{k}(\bF_{\rT}, \bW_{\rX,k})$ as the MSE matrix at relay $k \in \cK_{\rX}$, which is given by
\begin{eqnarray}
\bE_{k}(\bF_{\rT}, \bW_{\rX,k}) &= & \tr\left(\bW^{*}_{\rX,k}\left[\bH_{k,\mu(k)}\bF_{\rT,k}\bF^{*}_{\rT,k}\bH^{*}_{k,\mu(k)} + \bR_{\rX,k}(\bF_{\rT})\right]\bW_{\rX,k}\right)\nonumber\\
&& - \tr(\bW^{*}_{\rX,k}\bH_{k,\mu(k)}\bF_{\rT,k}) - \tr(\bF^{*}_{\rT,k}\bH^{*}_{k,\mu(k)}\bW_{\rX,k}) + d_{1,k}.\nonumber
\end{eqnarray}
The MSE of the estimate of $\bx_{1,k}$ based on $\bW^{*}_{\rX,k}\by_{\rX,k}$ is given by
\begin{eqnarray}
\varepsilon_{k}(\bF_{\rT}, \bW_{\rX,k}) &=& \bbE\|\bx_{1,k} - \bW^{*}_{\rX,k}\by_{\rX,k}\|^{2}_{\rF}\nonumber\\
&=& \tr(\bE_{k}(\bF_{\rT}, \bW_{\rX,k})).\label{eq:MSE}
\end{eqnarray}
Fixing $\bF_{\rT}$ and solving $\frac{\partial \bbE_{k}(\bF_{\rT}, \bW_{\rX,k})}{\partial \bW^{*}_{\rX,k}} = 0$, we can check that $\varepsilon_{k}(\bF_{\rT}, \bW_{\rX,k})$ is minimized by the linear MMSE receive filter $\bW^{\rMMSE}_{\rX,k}$ given in (\ref{eq:MMSEreceiveFilter}). We denote $\bE_{k,0}(\bF_{\rT}) = \bE_{k}(\bF_{\rT}, \bW^{\rMMSE}_{\rX,k})$. After some manipulation, we obtain the following well-known relationship between $R_{1,k}(\bF_{\rT})$ and $\bE_{k,0}(\bF_{\rT})$
\begin{equation}
R_{1,k}(\bF_{\rT}) = -\log_{2}(\det(\bE_{k,0}(\bF_{\rT}))),
\end{equation}
or equivalently
\begin{equation}\label{eq:relationship}
\xi_{1,k}(\bF_{\rT}) = 1/\det(\bE_{k,0}(\bF_{\rT})) - 1.
\end{equation}

Based on (\ref{eq:relationship}), we define
\begin{eqnarray}
g_{k}(\bE_{k,0}(\bF_{\rT})) 
&\triangleq& - \frac{\log 2}{t} u_{k}\left(1/\det(\bE_{k,0}(\bF_{\rT})) - 1, t, \bar{\xi}_{2,k}\right)\\
&=& \begin{cases}
\log\left(\det(\bE_{k,0}(\bF_{\rT}))\right), \mbox{~if~} \det(\bE_{k,0}(\bF_{\rT})) \geq (1 + \eta_{k})^{-1},\\
-\log(1 + \eta_{k}) - \exp\left[1 - (1+\eta_{k})\det(\bE_{k,0}(\bF_{\rT}))\right] + 1, \mbox{~otherwise}.
\end{cases}\label{eq:gk}
\end{eqnarray}
We can check that $g_{k}(\bE_{k,0}(\bF_{\rT}))$ is twice continuously differentiable with respect to $\bE_{k,0}(\bF_{\rT})$ at any point. Moreover, $g_{k}(\bE_{k,0}(\bF_{\rT}))$ is a strictly concave function of $\bE_{k,0}(\bF_{\rT})$. Note that $(\cA\cF\cP)$ is equivalent to the following optimization problem
\begin{eqnarray}
(\cG\cF\cP):~\min_{\bF_{\rT} \in \bbF_{\rT}} &&\sum_{k\in\cK_{\rX}} g_{k}(\bE_{k,0}(\bF_{\rT}))\nonumber\\
\mbox{subject~to} && \sum_{\substack{m\in \cK_{\rX}\\\mu(m) = k}} \tr(\bF_{\rT,m}\bF^{*}_{\rT,m}) \leq p_{\rT,k}, \forall k \in \cK_{\rT}.
\end{eqnarray}
Note that $(\cG\cF\cP)$ is nonconvex and NP-hard. We define the following function
\begin{eqnarray}
\alpha(x) = 
\begin{cases}
1, \mbox{~if~} x \geq (1 + \eta_{k})^{-1}\\
(1+\eta_{k})x\exp\left[(1+\eta_{k})x - 1\right], \mbox{~otherwise}.
\end{cases}
\end{eqnarray}
The first-order gradient of $g_{k}(\bE_{k,0}(\bF_{\rT}))$ with respect to $\bE_{k,0}(\bF_{\rT})$ is given by
\begin{eqnarray}
\nabla g_{k}(\bE_{k,0}(\bF_{\rT})) &\triangleq& \frac{\partial g_{k}(\bE_{k,0}(\bF_{\rT}))}{\partial \bE_{k,0}(\bF_{\rT})}\nonumber\\
&=& \alpha(\det(\bE_{k,0}(\bF_{\rT})))(\bE_{k,0}(\bF_{\rT}))^{-1}.\label{eq:gGradient}
\end{eqnarray}
According to Theorem 2 in~\cite{ShiEtAl2011:TSP}, the inverse mapping of $\nabla g_{k}(\bE_{k,0}(\bF_{\rT}))$ is well-defined. We refer to it as $\gamma_{k}(\cdot): \bbC^{d_{1,k} \times d_{1,k}}\rightarrow \bbC^{d_{1,k} \times d_{1,k}}$.

We now use the technique in~\cite{ShiEtAl2011:TSP,RazaviyaynEtAl2011:CISS} to solve $(\cA\cF\cP)$ via matrix-weighted sum-MSE minimization. We introduce auxiliary variables $\bV_{k} \in \bbC^{d_{1,k} \times d_{1,k}}$ for $k \in \cK_{\rX}$. We define $\bV \triangleq (\bV_{1},\cdots, \bV_{K_{\rX}}) \in \bbV \triangleq \bbC^{d_{1,1} \times d_{1,1}} \times \cdots \times \bbC^{d_{1,K_{\rX}} \times d_{1,K_{\rX}}}$. We define the following matrix-weighted sum-MSE function
\begin{equation}
s_{k}(\bF_{\rT}, \bW_{\rX}, \bV) = \tr(\bV^{*}_{k}\bE_{k}(\bF_{\rT}, \bW_{\rX,k})) + g_{k}(\gamma_{k}(\bV_{k}))  - \tr(\bV^{*}_{k})\gamma_{k}(\bV_{k}).
\end{equation}
A matrix-weighted sum-MSE minimization problem is formulated as follows
\begin{eqnarray}
(\cM\cF\cP):~\min_{(\bF_{\rT},\bW_{\rX},\bV) \in \bbF_{\rT} \times \bbW_{\rX} \times \bbV} && \sum_{k \in \cK_{\rX}} s_{k}(\bF_{\rT}, \bW_{\rX}, \bV)\nonumber\\
\mbox{subject~to} && \sum_{\substack{m\in \cK_{\rX}\\\mu(m) = k}} \tr(\bF_{\rT,m}\bF^{*}_{\rT,m}) \leq p_{\rT,k}, \forall k \in \cK_{\rT}.\nonumber
\end{eqnarray}
It follows from Theorem 2 in~\cite{ShiEtAl2011:TSP} that $(\cM\cF\cP)$ and $(\cG\cF\cP)$ have the same stationary points if the relays use their corresponding linear MMSE receive filters and the matrix weights $\bV_{k} = \nabla g_{k}(\bE_{k}(\bF_{\rT},\bW_{\rX}))$ for any $\bF_{\rT}$ and $\bW_{\rX}$. Thus, we can find the stationary points of $(\cA\cF\cP)$ by solving $(\cG\cF\cP)$.

\subsubsection{Algorithm for Matrix-Weighted Sum-MSE Minimization}
We adopt an alternating minimization approach to develop an iterative algorithm for finding the stationary points of $(\cM\cF\cP)$. In each iteration, we focus on determining only one of the sets of parameters $\bF_{\rT}, \bW_{\rX}$, and $\bV$ while assuming the others are fixed. When $\bF_{\rX}$ and $\bV$ are fixed, the optimal linear receive filter at relay $k \in \cK_{\rX}$ is exactly $\bW^{\rMMSE}_{\rX,k}$ given in (\ref{eq:MMSEreceiveFilter}). In addition, as discussed in Section \ref{subsubsec:sumMMSE}, when $\bF_{\rT}$ and $\bW_{\rX}$ are fixed, the optimal matrix weights are as follows 
\begin{equation}
\bV^{\ropt}_{k} = \nabla g_{k}(\bE_{k,0}(\bF_{\rT})), \forall k \in \cK_{\rX}.\label{eq:Vopt}
\end{equation}
It follows from (\ref{eq:MSE}) that $\bE_{k}(\bF_{\rT}, \bW_{\rX,k})$ is a Hermitian and positive semidefinite matrix for any $\bF_{\rT}$ and $\bW_{\rX}$. Combined with (\ref{eq:gGradient}), we have $\nabla g_{k}(\bE_{k,0}(\bF_{\rT}))$ is a Hermitian and positive semidefinite matrix. This means that if we always choose $\bV_{k} = \bV^{\ropt}_{k}$ according to (\ref{eq:Vopt}), then $\bV_{k}$ is a Hermitian and positive semidefinite matrix for $k \in \cK_{\rX}$. 

What remains is the design of $\bF_{\rT}$ when $\bW_{\rX}$ and $\bV_{\rX}$ are fixed. When $\bW_{\rX}$ and $\bV_{\rX}$ are fixed, we need to solve the following optimization problem to determine $\bF_{\rT}$
\begin{eqnarray}
(\cM\cF\cP\mbox{-}\cF_{\rT}):~\min_{\bF_{\rT} \in \bbF_{\rT}} && \sum_{k \in \cK_{\rX}} \tr(\bV^{*}_{k}\bE_{k}(\bF_{\rT},\bW_{\rX,k}))\nonumber\\
\mbox{subject~to} && \sum_{\substack{m\in \cK_{\rX}\\\mu(m) = k}} \tr(\bF_{\rT,m}\bF^{*}_{\rT,m}) \leq p_{\rT,k}, \forall k \in \cK_{\rT}.
\end{eqnarray}
Note that $(\cM\cF\cP\mbox{-}\cF_{\rT})$ is convex with respect to $\bF_{\rT,k}$ for $k \in \cK_{\rX}$. Let $\lambda_{k} \geq 0$ be the Lagrangian multiplier associated with the sum transmit power constraint at transmitter $k \in \cK_{\rT}$. Based on the optimality condition of $(\cM\cF\cP\mbox{-}\cF_{\rT})$, the globally optimal solution of $(\cM\cF\cP\mbox{-}\cF_{\rT})$ must has the following form for $k \in \cK_{\rX}$
\begin{eqnarray}\label{eq:Fopt}
\bF_{\rT,k}(\lambda_{k}) &=&\left(\sum_{m\in\cK_{\rX}} \bH^{*}_{m,\mu(k)}\bW_{\rX,m}\bV_{m}\bW^{*}_{\rX,m}\bH^{*}_{m,\mu(k)} + \lambda_{k}\bI_{N_{\rT,k}}\right)^{-1}\bH^{*}_{k,\mu(k)}\bW_{\rX,k}\bV_{k}.
\end{eqnarray}
Since $\bV_{k}$ is a Hermitian and positive semidefinite matrix, $\sum_{m\in\cK_{\rX}} \bH^{*}_{m,\mu(k)}\bW_{\rX,m}\bV_{m}\bW^{*}_{\rX,m}\bH^{*}_{m,\mu(k)}$ is also Hermitian and positive semidefinite. It can be checked that $\tr\big(\bF_{\rT,k}(\lambda_{k})\bF^{*}_{\rT,k}(\lambda_{k})\big)$ is strictly decreasing with $\lambda_{k}$ in $[0,+\infty)$. The optimal Lagrangian multiplier $\lambda^{*}_{k} \geq 0$ is chosen such that the complementary slackness condition of the sum power constraint at transmitter $k \in \cK_{\rT}$ is satisfied. For any $k \in \cK_{\rT}$, if $\sum_{m\in\cK_{\rX}:~\mu(m) = k} \tr(\bF_{\rT,k}(0)\bF^{*}_{\rT,k}(0)) \leq p_{\rT,k}$, then $\bF^{\ropt}_{\rT,k} = \bF_{\rT,k}(0)$. Otherwise, $\lambda^{*}_{k}$ is the unique solution of the following equation 
\begin{equation}
\sum_{\substack{m \in \cK_{\rX}\\\mu(m) = k}} \tr\Big(\bF_{\rT,k}(\lambda_{k})\bF^{*}_{\rT,k}(\lambda_{k})\Big) = p_{\rT,k}.
\end{equation}
This equation can be solved by using one-dimensional search techniques, e.g., the bisection method.

Note that in the proposed algorithm for solving $(\cM\cF\cP)$, we are able to find the globally optimal solutions to the corresponding optimization problem in each iteration. Therefore, the algorithm is guaranteed to converge to a stationary point of $(\cM\cF\cP)$, which is also a stationary point of $(\cA\cF\cP)$. Let  $\bar{\bF}_{\rT}$ denote the transmit precoders corresponding to the resulting stationary point. It is worthwhile to emphasize that it is not guaranteed that we can find a stationary point of $(\cF\cP)$. Thus, a two-hop rate mismatch may still happen for the resulting suboptimal solution $(\bar{\bF}_{\rT}, \bar{\bF}_{\rX})$ of the original transmit precoder design problem $(\cOP)$. This leaves room for potential improvements in terms of maximizing $R_{\rsum}(\bF_{\rT},\bar{\bF}_{\rX})$.

\subsubsection{Rate-Matching Transmit Power Control}\label{subsubsec:PC}
We propose an iterative power control method for eliminating any residual two-hop rate mismatch corresponding to $(\bar{\bF}_{\rT}, \bar{\bF}_{\rX})$. Let $\bF^{(n)}_{\rT, k}$ denote the transmit precoder for the transmission to relay $k\in \cK_{\rX}$ in iteration $n$ of this method. Note that $\bF^{(0)}_{\rT,k} = \bar{\bF}_{\rT,k}$ for $k \in \cK_{\rX}$. Let $\theta^{(n)}_{k} \in \bbR$ be the norm of $\bF^{(n)}_{\rT,k}$, i.e., the power allocated for the transmission from transmitter $\mu(k) \in \cK_{\rT}$ to relay $k \in \cK_{\rX}$ in iteration $n$. We propose to fix the shapes of the transmit precoders and to adjust only their norm $\theta_{k}$ for $k \in \cK_{\rX}$. It follows that $\bF^{(n)}_{\rT,k} = \theta^{(n)}_{k} \frac{\bar{\bF}_{\rT,k}}{\|\bar{\bF}_{\rT,k}\|_{\rF}}$ for all $n\geq 0$ where $\frac{\bar{\bF}_{\rT,k}}{\|\bar{\bF}_{\rT,k}\|_{\rF}}$ has unit norm and represents the shape of $\bar{\bF}_{\rT,k}$. 

We define $\boldsymbol{\theta}^{(n)} = (\theta^{(n)}_{1}, \cdots, \theta^{(n)}_{K_{\rX}}) \in \bbR^{K_{\rX} \times 1}$ and $\boldsymbol{\theta}^{(n)}_{-k} = (\theta_{1}^{(n)}, \cdots, \theta_{k-1}^{(n)}, \theta_{k+1}^{(n)}\cdots, \theta_{K_{\rX}}^{(n)})$. Note that $\boldsymbol{\theta}^{(n)}$ and $(\theta_{k}^{(n)}; \boldsymbol{\theta}_{-k}^{(n)})$ are used interchangeably in the section. Also, we denote $\bcH_{m,\mu(k)} = \bH_{m,\mu(k)}\frac{\bar{\bF}_{\rT,k}}{\|\bar{\bF}_{\rT,k}\|_{\rF}}$  for $k, m \in \cK_{\rX}$. The interference plus noise covariance matrix at relay $k \in \cK_{\rX}$ is rewritten as
\begin{equation}\label{eq:PCnoiseCovMat}
\bR_{\rX,k}(\boldsymbol{\theta}^{(n)}_{-k}) = \sum_{\substack{m \in\cK_{\rX}\\ m\neq k}} \theta^{(n)}_{m}\bcH_{k,\mu(m)}\bcH^{*}_{k,\mu(m)} + \sigma^{2}_{\rX,k}\bI_{N_{\rX,k}},
\end{equation}
while the maximum achievable rate at relay $k \in \cK_{\rX}$ is rewritten as
\begin{equation}
R_{1,k}(\boldsymbol{\theta}^{(n)}) = \log_{2}\det\Big(\bI + \theta^{(n)}_{k}\bcH^{*}_{k,\mu(k)}\big[\bR_{\rX,k}(\boldsymbol{\theta}^{(n)}_{-k})\big]^{-1}\bcH_{k,\mu(k)}\Big).
\end{equation}
Thus, it follows that
\begin{equation}\label{eq:PCxi1}
\xi_{1,k}(\boldsymbol{\theta}^{(n)}) = \det\Big(\bI + \theta^{(n)}_{k}\bcH^{*}_{k,\mu(k)}\big[\bR_{\rX,k}(\boldsymbol{\theta}^{(n)}_{-k})\big]^{-1}\bcH_{k,\mu(k)}\Big) - 1.
\end{equation}
The end-to-end achievable rate corresponding to relay $k \in \cK_{\rX}$ is written as
\begin{equation}
R_{k}(\boldsymbol{\theta}^{(n)}) = \min\{tR_{1,k}(\boldsymbol{\theta}^{(n)}), (1 - t)\log_{2}(1+\bar{\xi}_{2,k})\}.
\end{equation}
Note that $\bcH^{*}_{k,\mu{k}}[\bR_{\rX,k}(\boldsymbol{\theta}^{(n)}_{-k})]^{-1}\bcH_{k,\mu{k}}$ is independent of $\theta^{(n)}_{k}$. It is also a Hermitian and positive semidefinite matrix. Since $\det(\bI+x\bA)$ is strictly increasing in $x$ for $x \geq 0$ when $\bA$ is a positive semidefinite matrix, both $R_{1,k}(\boldsymbol{\theta}^{(n)})$ and $\xi_{1,k}(\boldsymbol{\theta}^{(n)})$ are strictly increasing in $\theta^{(n)}_{k}$ when $\theta^{(n)}_{k} \geq 0$. 

Let $\cA^{(n)} \triangleq \{k \in \cK_{\rX}:~tR_{1,k}(\boldsymbol{\theta}^{(n)}) > (1 - t)\log_{2}(1+\bar{\xi}_{2,k})\}$ be the index set of the relays with a dominant first hop in iteration $n$. If $k \in \cA^{(n)}$, then excess power is allocated for the transmission to relay $k$. Similarly, let $\cB^{(n)}  \triangleq \{k \in \cK_{\rX}:~tR_{1,k}(\boldsymbol{\theta}^{(n)}) < (1 - t)\log_{2}(1+\bar{\xi}_{2,k})\}$ be the index set of the relays with a dominant second hop in iteration $n$. A two-hop rate mismatch happens if and only if $\cA^{(n)} \not\equiv \emptyset$ and $\cB^{(n)} \not\equiv \emptyset$. When a two-hop rate mismatch happens, we consider an arbitrary $k_{\cA}\in \cA^{(n)}$ and $k_{\cB} \in \cB^{(n)}$. It follows $\cA^{(n)} \cap \cB^{(n)} \equiv \emptyset$ that $k_{\cA} \neq k_{\cB}$. When $\theta^{(n)}_{m}$ is fixed for $m \in \cK_{\rX}$ and $ m \neq k_{\cA}$, since $k_{\cA} \in \cA^{(n)}$, there must exist $\phi^{(n)}_{k_{\cA}} \in (0,\theta^{(n)}_{k_{\cA}})$ such that
\begin{equation}
tR_{1,k}(\phi^{(n)}_{k_{\cA}}; \boldsymbol{\theta}^{(n)}_{-k_{\cA}})) = (1-t) \log_{2}(1 + \bar{\xi}_{2,k_{\cA}}).
\end{equation}
Equivalently, it follows that
\begin{equation}\label{eq:PCphi}
\det\Big(\bI + \phi^{(n)}_{k_{\cA}}\bcH^{*}_{k_{\cA},\mu(k_{\cA})}\big[\bR_{\rX,k_{\cA}}(\boldsymbol{\theta}^{(n)}_{-k_{\cA}})\big]^{-1}\bcH_{k_{\cA},\mu(k_{\cA})}\Big) - 1 = \eta_{k_{\cA}}.
\end{equation}
Since the left-hand side of (\ref{eq:PCphi}) is strictly increasing in $\phi^{(n)}_{k_{\cA}}$, this equation has a unique solution, which can be found by using one dimensional search techniques, e.g., the bisection method. Note that $\phi^{(n)}_{k_{\cA}}$ is the rate-matching power for the transmission to relay $k_{\cA}$ in iteration $n$.

The key observation for the proposed power control algorithm is that if a two-hop rate mismatch happens at the end of iteration $n$, excess power can be reduced by setting $\theta^{(n+1)}_{k_{\cA}} = \phi_{k_{\cA}}$  and $\boldsymbol{\theta}^{(n+1)}_{-k_{\cA}} = \boldsymbol{\theta}^{(n)}_{-k_{\cA}}$ for an arbitrary $k_{\cA} \in \cA^{(n)}$. Note that $R_{k_{\cA}}(\boldsymbol{\theta}^{(n+1)}) = (1-t) \log_{2}(1 + \bar{\xi}_{2,k_{\cA}}) = R_{k_{\cA}}(\boldsymbol{\theta}^{(n)})$, Also, this excess power reduction decreases the power of interference observed by all relay $m \neq k_{\rA}$, leading to $R_{1,m}(\boldsymbol{\theta}^{(n+1)}) \geq R_{1,m}(\boldsymbol{\theta}^{(n)})$ and hence 
$R_{m}(\boldsymbol{\theta}^{(n+1)}) \geq R_{m}(\boldsymbol{\theta}^{(n)})$. Especially, it follows that $R_{1,k_{\cB}}(\boldsymbol{\theta}^{(n+1)}) > R_{1,k_{\cB}}(\boldsymbol{\theta}^{(n)})$ and hence $R_{k_{\cB}}(\boldsymbol{\theta}^{(n+1)}) > R_{k_{\cB}}(\boldsymbol{\theta}^{(n)})$. As a result, the end-to-end sum-rate is strictly improved, i.e., $\sum_{k\in\cK_{\rX}} R_{k}(\boldsymbol{\theta}^{(n)}) < \sum_{k\in\cK_{\rX}}R_{k}(\boldsymbol{\theta}^{(n+1)})$. Thus, when a two-hop rate mismatch happens, reducing excess power in a controlled manner strictly increases the end-to-end sum-rates.

Based on the observation, we propose an iterative algorithm for updating the power allocated for the transmission from the transmitters to the relays. Specifically, at the end of each iteration $n \geq 0$, each relay $k \in \cK_{\rR}$ computes $\xi_{1,k}(\boldsymbol{\theta}^{(n)})$ to check if the transmission to itself is allocated excess power. If $\xi_{1,k}(\boldsymbol{\theta}^{(n)}) \leq \eta_{k}$, then the power allocated for the transmission to relay $k$ does not need to change, i.e., $\theta^{(n+1)}_{k} = \theta^{(n)}_{k}$. Otherwise, relay $k$ determines the corresponding rate-matching power $\phi^{(n)}_{k}$ and feeds back the value to its associated transmitter $\mu(k)$ to instruct the transmitter to update $\theta^{(n+1)}_{k} = \phi^{(n+1)}_{k} \in (0,\theta^{(n)}_{k})$. In other words, the power update rule for $k \in \cK_{\rX}$ and $n \geq 0$ is
\begin{equation}\label{eq:PCupdateRule}
\theta^{(n+1)}_{k} = \begin{cases}
\phi^{(n)}_{k}, \mbox{~if~} k \in \cA^{(n)},\\
\theta^{(n)}_{k}, \mbox{~otherwise}.
\end{cases}
\end{equation}
This process is repeated until the algorithm converges or the maximum number of iteration is reached. It follows from the power update rule that $\theta^{(n+1)}_{k} \leq \theta_{k}^{(n)}$ for $k\in\cK_{\rX}$ and $n \geq 0$. Several properties of the proposed algorithm are presented in Remark \ref{rem:PConvergence}, Remark \ref{rem:PCnoRateMismatch}, and Remark \ref{rem:PCnondecreasingRates}.

\begin{Remark}\label{rem:PConvergence}
Note that $\theta^{(n+1)}_{k} \leq \theta^{(n)}_{k}$ for $k \in \cK_{\rX}$ and $n \geq 1$. Since $\theta^{(n)}_{k}$ is nonnegative, it is lower bounded by $0$. Thus, it is guaranteed that the proposed power control algorithm converge as the number of iterations goes to infinity.
\end{Remark}

\begin{Remark}\label{rem:PCnoRateMismatch}
The resulting solution of the proposed power control algorithm does not causes a two-hop rate mismatch. Indeed, let $n_{0}$ be the index of the iteration at which the proposed algorithm is convergent, i.e., $\theta^{(n_{0})}_{k} = \theta^{(n_{0} + 1)}_{k}$ for $k \in \cK_{\rX}$. This means that $\xi_{1,k}(\boldsymbol{\theta}^{(n_{0})}) \leq \eta_{k}$ for all $k \in \cK_{\rX}$ or $\cA^{(n_{0})} \equiv \emptyset$.  Recall that a two-hop rate mismatch happens in iteration $n$ if and only if $\cA^{(n)} \not\equiv \emptyset$ and  $\cB^{(n)} \not\equiv \emptyset$. Thus there is not  any two-hop rate mismatch in iteration $n_{0}$. 
\end{Remark}

\begin{Remark}\label{rem:PCnondecreasingRates}
The proposed algorithm does not decrease $R_{k}(\boldsymbol{\theta}^{(n)})$ over iterations for $n \geq 0$. It can be showed that if $\bA$ is positive semidefinite, then $\det(\bI + \bB^{*}(\bI + x\bA)^{-1}\bB)$ is strictly decreasing in $x$ for $x\geq 0$ for any $\bB$. From (\ref{eq:PCnoiseCovMat}) and (\ref{eq:PCxi1}), we have
\begin{eqnarray}
\xi_{1,k}(\boldsymbol{\theta}^{(n + 1)}) + 1
&=& \det\Bigg(\bI + \theta^{(n+1)}_{k}\bcH^{*}_{k,\mu(k)}\bigg[\sum_{\substack{m \in \cK_{\rX}\\ m\neq k}} \theta^{(n + 1)}_{m}\bcH_{k,\mu(m)}\bcH^{*}_{k,\mu(m)} + \sigma^{2}_{k}\bI\bigg]^{-1}\bcH_{k,\mu(k)}\Bigg) \nonumber\\
&\geq& \det\Bigg(\bI + \theta^{(n + 1)}_{k}\bcH^{*}_{k,\mu(k)}\bigg[\sum_{\substack{m \in \cK_{\rX}\\ m\neq k}} \theta^{(n)}_{m}\bcH_{k,\mu(m)}\bcH^{*}_{k,\mu(m)} + \sigma^{2}_{k}\bI\bigg]^{-1}\bcH_{k,\mu(k)}\Bigg)\label{eq:PC1}\\
&=&\begin{cases}
\eta_{k} + 1, \mbox{~if~} k\in \cA^{(n)}\\
\xi_{1,k}(\boldsymbol{\theta}^{(n)}) + 1, \mbox{~otherwise},
\end{cases}\label{eq:PC2}
\end{eqnarray}
where (\ref{eq:PC1}) comes from the property that $\theta^{(n+1)}_{k} \leq \theta^{(n)}_{k}$ for $k \in \cK_{\rX}$ and (\ref{eq:PC2}) comes from (\ref{eq:PCupdateRule}) and (\ref{eq:PCphi}). It follows that $\min\{\eta_{k}, \xi^{(n+1)}_{1,k}\} \geq \min\{\eta_{k}, \xi^{(n)}_{1,k}\}$; equivalently, $R_{k}(\boldsymbol{\theta}^{(n + 1)}) \geq R_{k}(\boldsymbol{\theta}^{(n)})$ for $k \in \cK_{\rX}$.
\end{Remark}

\begin{Remark}\label{rem:PCorderOptimization}
We can always use the same steps to develop a similar rate-matching transmit power control at the relays on the second hop. Nevertheless, because the proposed algorithm is implemented in a distributed manner, it is unclear how to determine when the rate-matching transmit power control should be performed on the first hop and when it should be performed on the second hop.
\end{Remark}

Note that the proposed power control algorithm does not help find any stationary points of $(\cOP)$. Essentially, it helps eliminate the residual two-hop rate mismatch in $(\bar{\bF}_{\rT}, \bar{\bF}_{\rX})$ to find a potentially better suboptimal solution to $(\cOP)$.

\subsection{Discussion}\label{subsec:discussion}

\subsubsection{Summary of the Proposed Algorithm}\label{subsubsec:summary}
\begin{figure}[!h]
\centering
\includegraphics[width=4.3in]{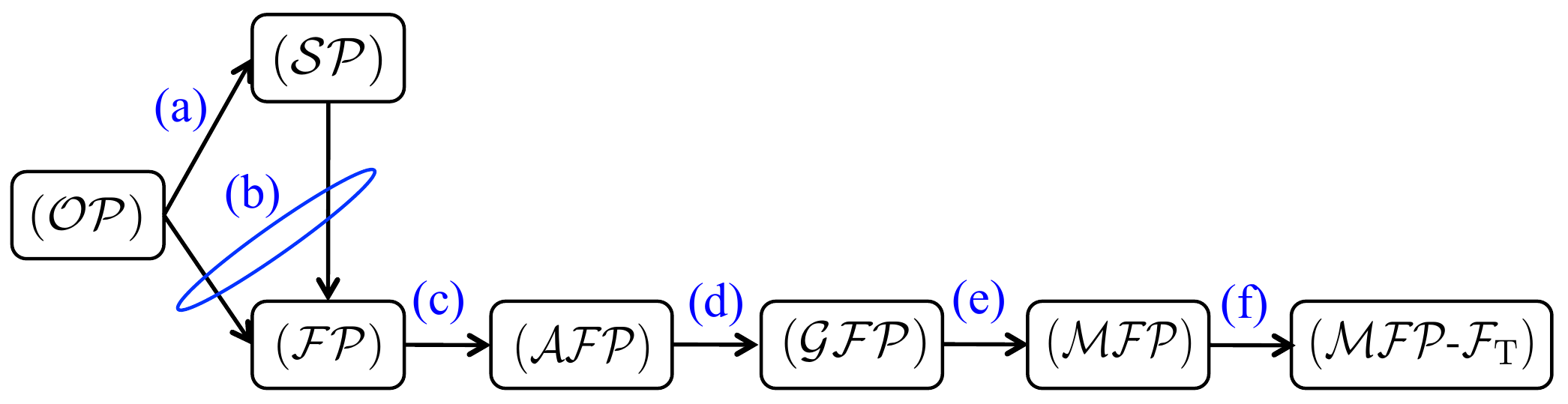}
\caption{Illustration of the transformations between different optimization problems to find high-quality suboptimal solutions to $(\cOP)$: (a) the design problem for the transmit precoders at the relays $(\cSP)$ is obtained from $(\cOP)$ when the first hop is ignored; (b) the design problem for the transmit precoders at the transmitters $(\cF\cP)$ is obtained from $(\cOP)$ and the first-hop performance results from the previous step; (c) approximates of the utility function of $(\cF\cP)$ given in (\ref{eq:uk}) are used to formulate $(\cA\cF\cP)$; (d) $(\cA\cF\cP)$ is equivalently rewritten as $(\cG\cF\cP)$; (e) $(\cM\cF\cP)$ is the matrix-weighted sum-MSE minimization problem that has the same stationary points as $(\cG\cF\cP)$; the alternating minimization approach is adopted to solve $(\cM\cF\cP)$ and (f) $(\cM\cF\cP\mbox{-}\cF_{\rT})$ is the design problem to design the transmit precoders at the transmitters in the corresponding iterations.}
\label{fig:problemTransformation}
\end{figure}

\begin{figure}[!h]
\centering
\includegraphics[width=4.3in]{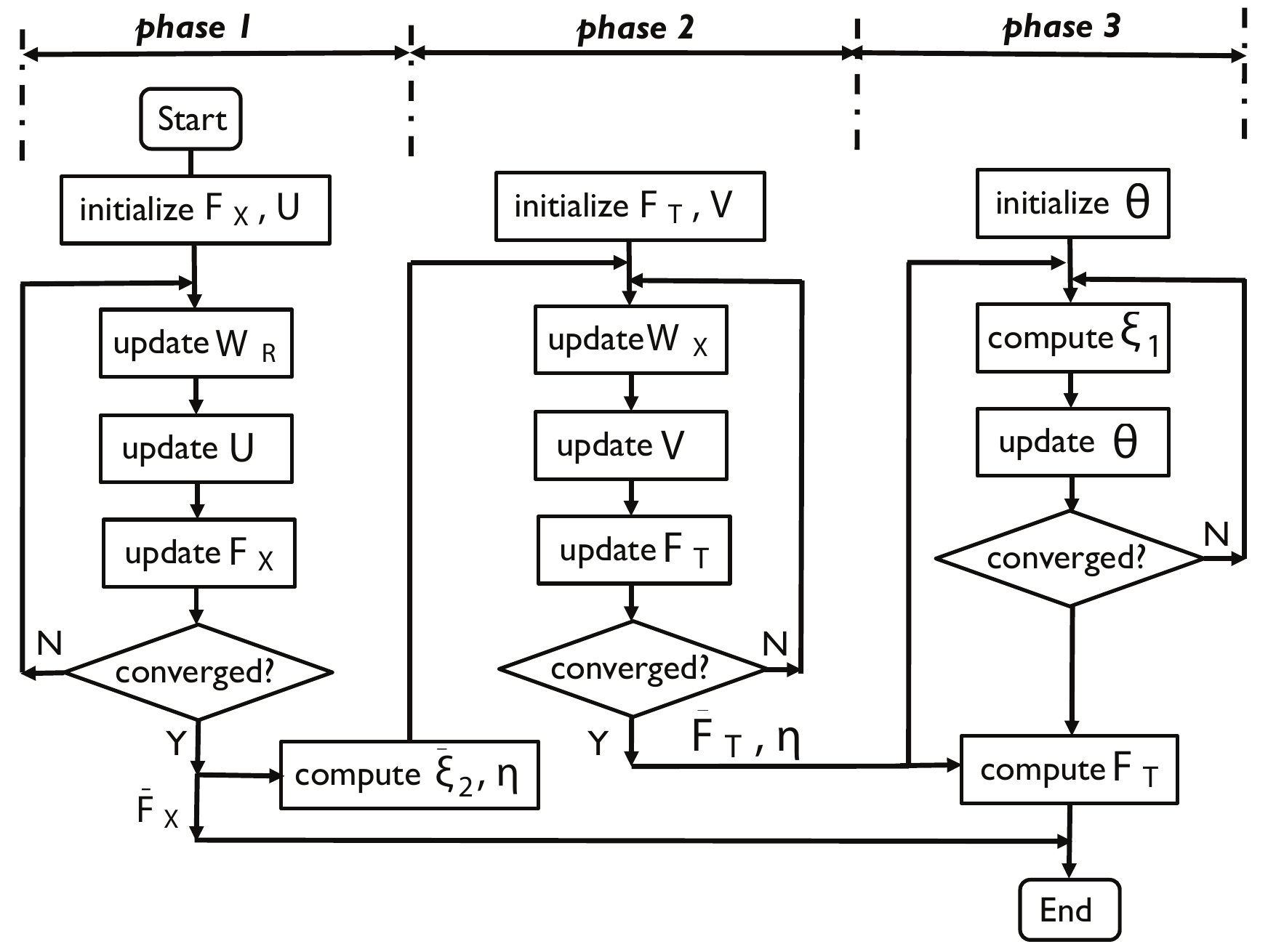}
\caption{Flow diagram of the proposed algorithm for the relay interference broadcast channel.}
\label{fig:relayIBCdiagram}
\vspace{-10pt}
\end{figure}

Recall that the proposed algorithm aims to find high-quality suboptimal solutions to the end-to-end sum-rate maximization problem $(\cOP)$. The development of this algorithm is involved with the formulation of a number of optimization problems as illustrated in Fig. \ref{fig:problemTransformation}. The flow diagram of the proposed algorithm is presented in Fig. \ref{fig:relayIBCdiagram}, where the notation for the main parameters is summarized in Table \ref{table:notation}. Note that Phase 1 is a counterpart of Table I in~\cite{ShiEtAl2011:TSP}.

\subsubsection{Distributed Implementation}\label{subsubsec:Implementation}
The proposed transmit precoding algorithm for the relay interference broadcast channel allows for distributed implementation. Similar to~\cite{RazaviyaynEtAl2011:CISS,ShiEtAl2011:TSP,ShiEtAl2009:ICC,ShiEtAl2009:MILCOM}, two assumptions are needed. First, each transmitting node has the corresponding local channel state information (CSI). Specifically, on the first hop, transmitter $k \in \cK_{\rT}$ has the CSI of $\bH_{m,k}$ for all $m \in \cK_{\rX}$; on the second hop, relay $m \in \cK_{\rX}$ has the CSI of $\bG_{q,m}$ for all $q \in \cK_{\rR}$. Second, there is a feedback channel to send information from a receiving node to its serving node, i.e., from receiver $q \in \cK_{\rR}$ to relay $\chi(q) \in \cK_{\rX}$ and from relay $m \in \cK_{\rX}$ to transmitter $\mu(m) \in \cK_{\rT}$. Specifically, while using the iteratively weighted MMSE approach to solve $(\cSP)$, in each iteration, the receivers need to feedback the updated matrix weights and receive filters to the relays~\cite{RazaviyaynEtAl2011:CISS,ShiEtAl2011:TSP}. At the end of Phase 1, each receiver $m \in \cK_{\rR}$ computes the second-hop received SINR and sends back to its serving relay, i.e. relay $\chi(m)$. On the first hop, in each iteration of solving $(\cM\cF\cP)$, the relays also need to send back the updated matrix weights and receive filters to the transmitters. 

\begin{table}[!h]
\centering
\caption{Summary of notation}\label{table:notation}
\begin{tabular}{| l | p{13cm} |}
\hline
Notation & Parameters\\
\hline
$\chi(k)$ & index of the relay aiding receiver $k \in\cK_{\rR}$\\
\hline
$\mu(k)$ & index of the transmitter aided by relay $k \in\cK_{\rX}$\\
\hline
$\bF_{\rX,q}$ & precoder at relay $\chi(q) \in \cK_{\rX}$ for transmission to receiver $q \in \cK_{\rR}$\\
\hline
$\bar{\bF}_{\rX,q}$ & resulting transmit precoder at relay $\chi(q) \in \cK_{\rX}$ for transmission to receiver $q \in \cK_{\rR}$\\
\hline
$\bW_{\rR,q}$ & receive filter at receiver $q \in \cK_{\rR}$\\
\hline
$\bU_{q}$ & matrix weight for MSE at receiver $q \in \cK_{\rR}$ in the second-hop weighted sum-MSEs\\
\hline
$\bar{\xi}_{2,k}$ & effective SINR corresponding to second-hop sum-rates from relay $k \in \cK_{\rX}$\\
\hline
$\eta_{k}$ & rate-matching SINR at relay $k \in \cK_{\rX}$ on the first hop\\
\hline
$\bF_{\rT,k}$ & precoder at transmitter $\mu(k) \in \cK_{\rT}$ for transmission to relay $k \in \cK_{\rX}$\\
\hline
$\bar{\bF}_{\rT,k}$ & resulting precoder at transmitter $\mu(k) \in \cK_{\rT}$ for transmission to relay $k \in \cK_{\rX}$ in the second phase\\
\hline
$\bW_{\rX,k}$ & receive filter at relay $k \in \cK_{\rX}$\\
\hline
$\bV_{k}$ & matrix weight for MSE at relay $k \in \cK_{\rX}$ in the first-hop weighted sum-MSEs\\
\hline
$\theta_{k}$ & Frobenius norm of $\bF_{\rT,k}$ in the third phase, i.e., the power for transmission from transmitter $\mu(k) \in \cK_{\rT}$ to relay $k \in \cK_{\rX}$\\
\hline
$\xi_{1,k}$ & effective received SINR at relay $k \in \cK_{\rX}$\\
\hline
\end{tabular}
\end{table}
 
\subsubsection{Opportunistic Approach}\label{subsubsec:opportunistic}

Recall that it is not guaranteed that the proposed algorithm can find optimal solutions to $(\cOP)$. In fact, the end-to-end sum-rate performance of the resulting solution $(\bF_{\rT}, \bF_{\rX})$ depends on the initial solutions in the first two phases. One method for improving the end-to-end sum-rate performance of the algorithm is to use multiple random transmit precoders as initial solutions in the first two phases and then to select the best one in terms of end-to-end sum-rate maximization. Such an opportunistic approach, however, may require more coordination among the nodes and result in more overhead in the network. 

\subsubsection{Order of Optimization}\label{subsubsec:optimizationOrder}
In the proposed algorithm, the relays are optimized for second-hop sum-rate maximization before the transmitters are designed for end-to-end sum-rate maximization. Note that the rates on two hops are matched with each other per relay in the design of the transmitters. In particular, when the allocated first-hop rates $\beta_{k,m}$ for $k \in \cK_{\rX}$ and $m \in \cK_{\rR}$ are determined as in (\ref{eq:beta}), at each relay $k$, the normalized first-hop rate $tR_{1,k}(\bF_{\rT})$ is matched with the sum of normalized second-hop rates $(1-t)R_{2,k,\rsum}(\bF_{\rX})$. There is another order of optimization where the transmitters are optimized for first-hop sum-rate maximization before the relays are optimized for end-to-end sum-rate maximization. In this case, the second-hop achievable rates $R_{2,m}(\bF_{\rX})$ for $m\in\cK_{\rR}$ are determined given the knowledge of the first-hop achievable rates $R_{1,k}(\bF_{\rT})$ for $k \in \cK_{\rX}$. For a fixed first-hop rate allocation strategy $\beta_{k,m}$ for $k \in \cK_{\rX}$ and $m \in \cK_{\rR}$, the end-to-end sum-rates in (\ref{eq:RsumDef}) can be rewritten as 
\begin{eqnarray}
R_{2,k, \rsum}(\bF_{\rX}) = \sum_{m \in \cK_{\rX}} \min\{\beta_{\chi(m),m},(1-t)R_{2,m}(\bF_{\rX})\}.\label{eq:RsumDef2}
\end{eqnarray}
Note that the end-to-end sum-rate performance in this case depends significantly on the selected first-hop rate allocation strategy. Moreover, when the set of $\beta_{k,m}$ for $k \in \cK_{\rX}$ and $m \in \cK_{\rR}$ is given, the same steps as in Section \ref{subsec:BF1} can be used in the design of relays in this case if matching the rates on two hops is performed per receiver. What remains is to determine the optimal first-hop rate allocation strategy in terms of end-to-end sum-rate maximization. Unfortunately, it is challenging to do this before designing the relays, i.e., without the knowledge of second-hop achievable rates $R_{2,m}(\bF_{\rX})$ for $m \in \cK_{\rR}$. This is the main reason that we decided that the relays are designed before the transmitters. In the special case where each relay serves a single receiver, the two orders of optimization are the same since there is no need for first-hop rate allocation.

\section{Simulations}\label{sec:simulation}
This section presents Monte Carlo simulations to investigate the end-to-end sum-rate performance of the proposed algorithm. We consider only symmetric relay systems with $N_{\rT,k} = N_{\rT}$ and $p_{\rT,k} = p_{\rT}$ for $k \in \cK_{\rT}$; $N_{\rX,m}=N_{\rX}$, $d_{1,m} = d_{1}$, and $p_{\rX,m} = p_{\rX}$ for $m \in \cK_{\rX}$; and $N_{\rR,q} = N_{\rR}$ and $d_{2,q} = d_{2}$ for $q \in \cK_{\rR}$. Each transmitter is aided by $K_{\rX} / K_{\rT}$ relays. Each relay forwards data from its associated transmitter to $K_{\rR} / K_{\rX}$ receivers. We denote the system as $(N_{\rT}^{K_{\rT}} \times N_{\rX}^{K_{\rX}} \times N_{\rR}^{K_{\rR}}, d_{1} \times d_{2})$.  Except when it is stated explicitly, we use $t = 0.5$. The channels are flat both in time and in frequency. The channel coefficients on two hops are generated as i.i.d. zero-mean unit-variance complex Gaussian random variables. Path loss is not considered in the simulations, thus the average power values of all cross-links on the same hop are equal to each other. The power values are normalized such that $p_{\rT} = p_{\rX}$, $\sigma_{\rX,m} =1$ for $m \in \cK_{\rX}$ and $\sigma_{\rR,q} = 1$ for $q \in \cK_{\rR}$. The plots are produced by averaging over 1000 random channel realizations. In each channel realization, the initial transceivers are chosen randomly. The maximum number of iterations in the first two phases is 2000 while that of the last phase is 30.

\subsection{Convergence of the Proposed Algorithm}
\begin{figure}[!h]
\centering
\includegraphics[width=3.4in]{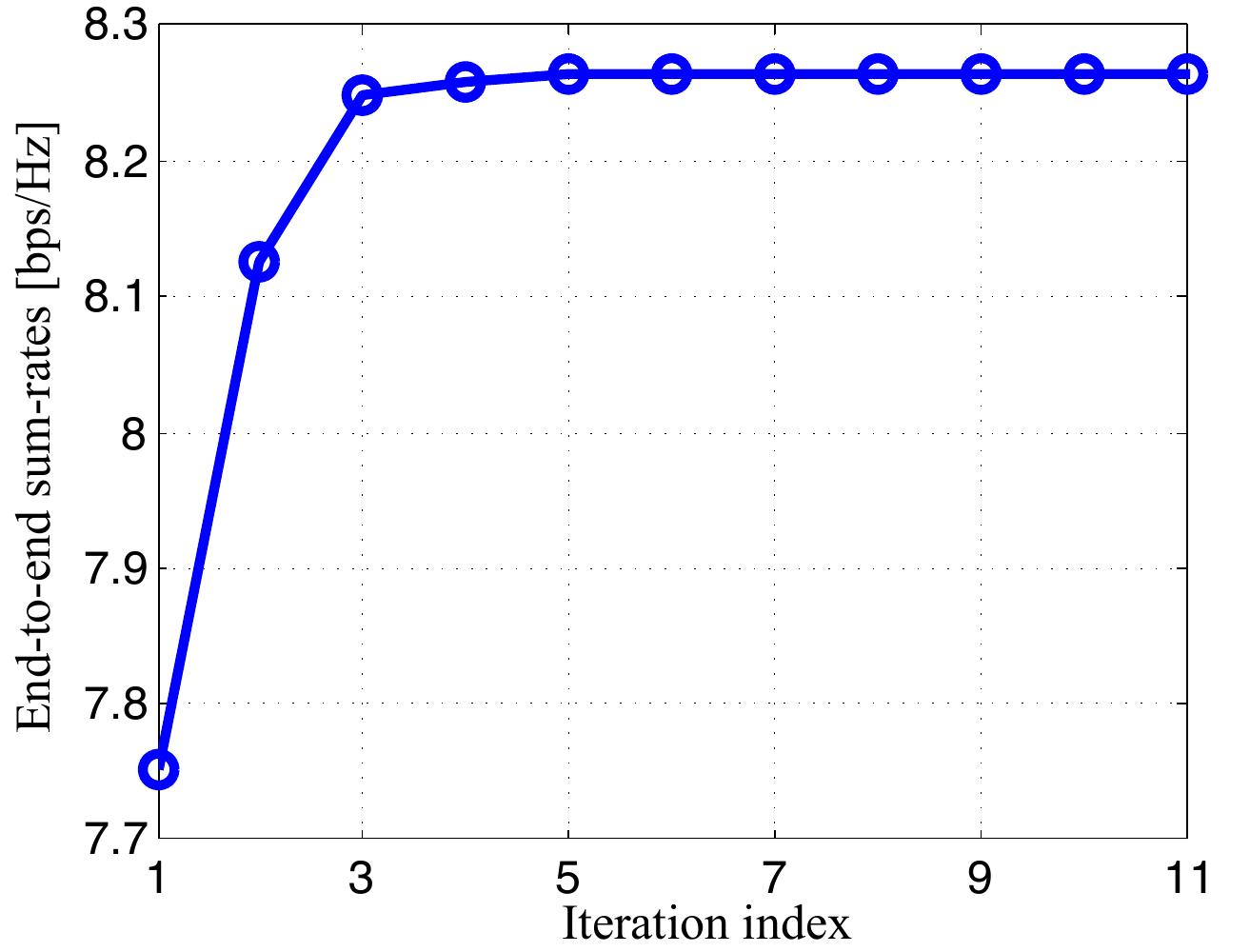}
\caption{Convergence behavior of the rate-matching power control algorithm for the $(2^{3}\times 2^{6}\times 2^{12}, 1 \times 1)$ system.}
\label{fig:convergencePCConf2B}
\end{figure}

Note that if each of its three phases is convergent, the proposed algorithm is also convergent. The first two phases are based on the prior work for the single-hop MIMO broadcast channel, of which the convergence is well validated in~\cite{ShiEtAl2011:TSP}. Fig. \ref{fig:convergencePCConf2B} shows the convergence behavior of the proposed rate-matching transmit power control method for a channel realization of the system $(2^{3}\times 2^{6}\times 2^{12}, 1 \times 1)$. We observe that this method converges in few iterations and it does so monotonically.

\subsection{Benefits of Each of the Last Two Phases}
\begin{figure}[!h]
\centering
\includegraphics[width=3.4in]{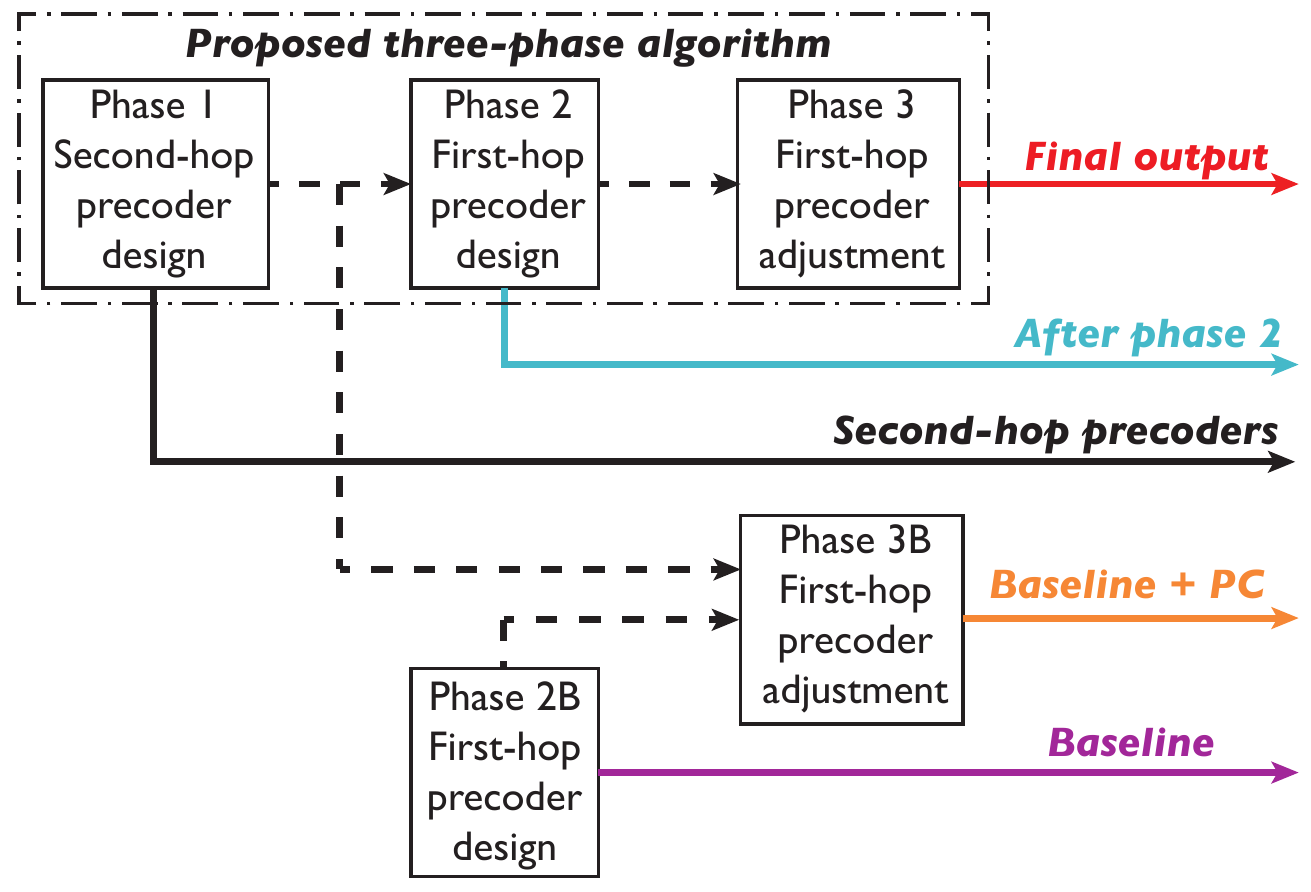}
\caption{Block diagrams of the proposed algorithm and the three reference algorithms. The dashed lines represent the links used for exchanging design parameters. The solid lines represent the output of the corresponding blocks.}
\label{fig:rateMatchingAlternatives}
\vspace{-10pt}
\end{figure}

In this section, we investigate the benefits of each of the last two phases by comparing the proposed algorithm with three alternative algorithms. Fig. \ref{fig:rateMatchingAlternatives} shows the block diagrams of all four algorithms. All algorithms use the same design of the second-hop precoders, which is the first phase of the proposed algorithm. Although they use different approaches to designing the first-hop precoders, they always use the same initial solution. The na\"{i}ve approach is used as the main baseline strategy, which is labeled as \emph{Baseline}. The output of the proposed algorithm is labeled as \emph{Final output}. The output of the second phase in the proposed algorithm is labeled as \emph{After phase 2}. Since the proposed rate-matching power control method for first-hop precoder adjustment can be applied for any $(\bF_{\rT}, \bF_{\rX})$, we proposed an algorithm that applies the proposed rate-matching power control for first-hop precoder adjustment to the output of Baseline. We label it as \emph{Baseline + PC}. Note that the comparison between \emph{After phase 2} and \emph{Baseline} gives us insights into the benefits of the second phase of the proposed algorithm. Similarly, the comparison between \emph{Final output} and \emph{After phase 2} and that between \emph{Baseline + PC} and \emph{Baseline} give us insights into the benefits of the third phase of the proposed algorithm. 

\begin{figure}[!h]
\centering
\includegraphics[width=3.4in]{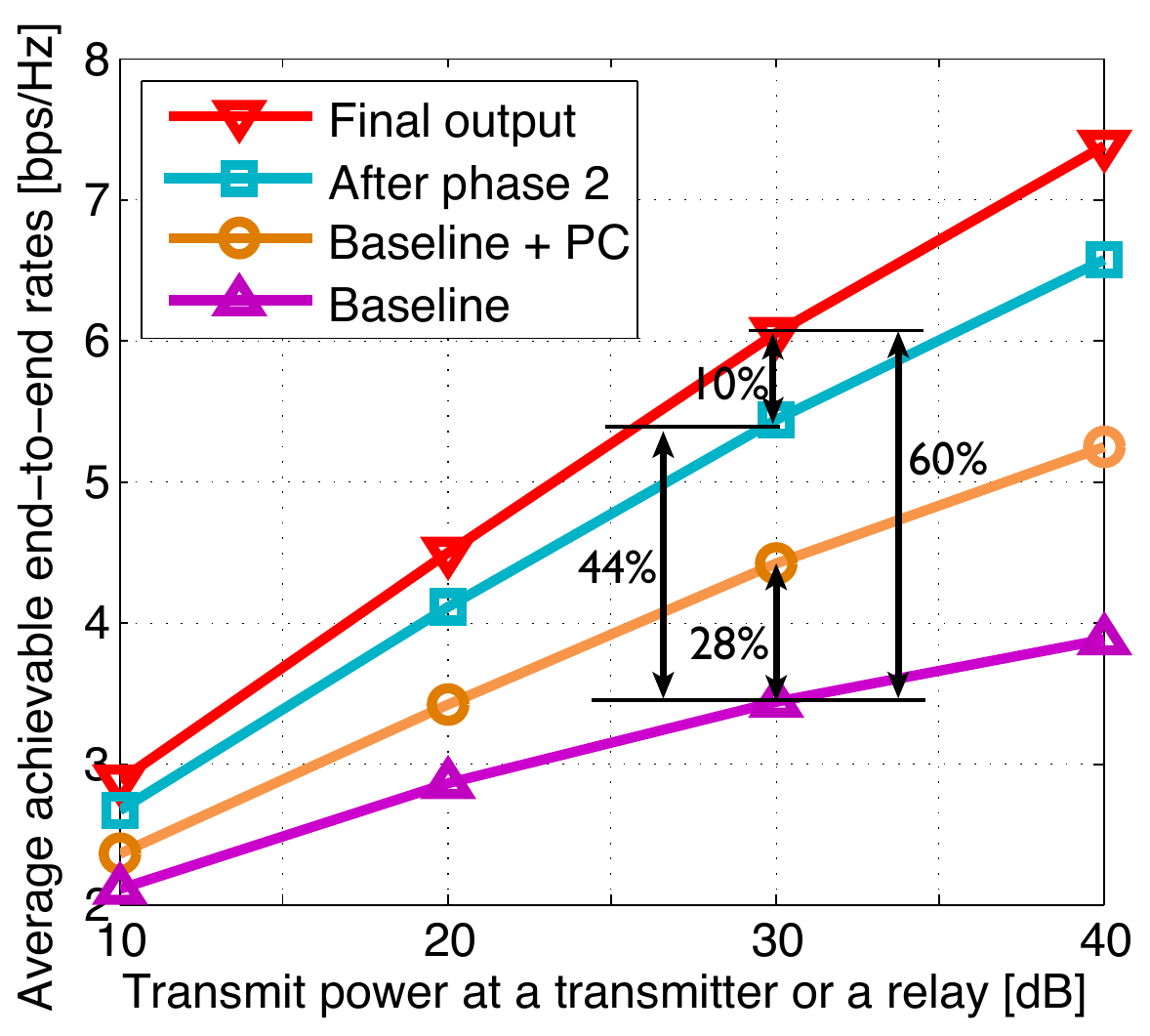}
\caption{Investigation of the benefits of the last two phases for the $(2^{3} \times 2^{6} \times 2^{12}, 1 \times 1)$ system.}
\label{fig:rateMatchingImpactConf2B}
\end{figure}

Fig. \ref{fig:rateMatchingImpactConf2B} shows the average end-to-end sum-rates achieved by the four algorithms for the $(2^{3} \times 2^{6} \times 2^{12}, 1 \times 1)$ system. All algorithms achieve non-zero end-to-end multiplexing gains thanks to their capabilities of interference management on two hops. The proposed algorithm outperforms the other algorithms for all transmit power values. At $p_{\rT} = p_{\rX} = 30$ dB, \emph{Final output} provides a gain of 10$\%$ over \emph{After phase 2}, which in turn provides a gain of 44$\%$ over \emph{Baseline}. Also, \emph{Baseline + PC} provides a gain of 28$\%$ over \emph{Baseline}. The results mean that thanks to the consideration of $t$ and $\xi_{2,k}$ for $k \in \cK_{\rX}$ in the design of $\bF_{\rT}$, the second phase of the proposed algorithm is able to alleviate a two-hop rate mismatch to obtain higher end-to-end sum-rates. In addition, the results show that there exist two-hop rate mismatches at the output of both \emph{After phase 2} and \emph{Baseline}. This is expected since the second phase of the proposed algorithm aims only at maximizing an approximate of the end-to-end sum-rates. We notice that the two-hop rate-mismatch of the output of \emph{Baseline} is relatively more significant than that of the output of \emph{After phase 2}. This emphasizes the benefits of two-hop rate-mismatch alleviation of the second phase of the proposed algorithm. Note that  \emph{After phase 2} outperforms \emph{Baseline + PC} in this simulation scenario, however, it is not guaranteed that \emph{After phase 2} always outperforms \emph{Baseline + PC}.

\begin{figure}[!h]
\centering
\includegraphics[width=3.4in]{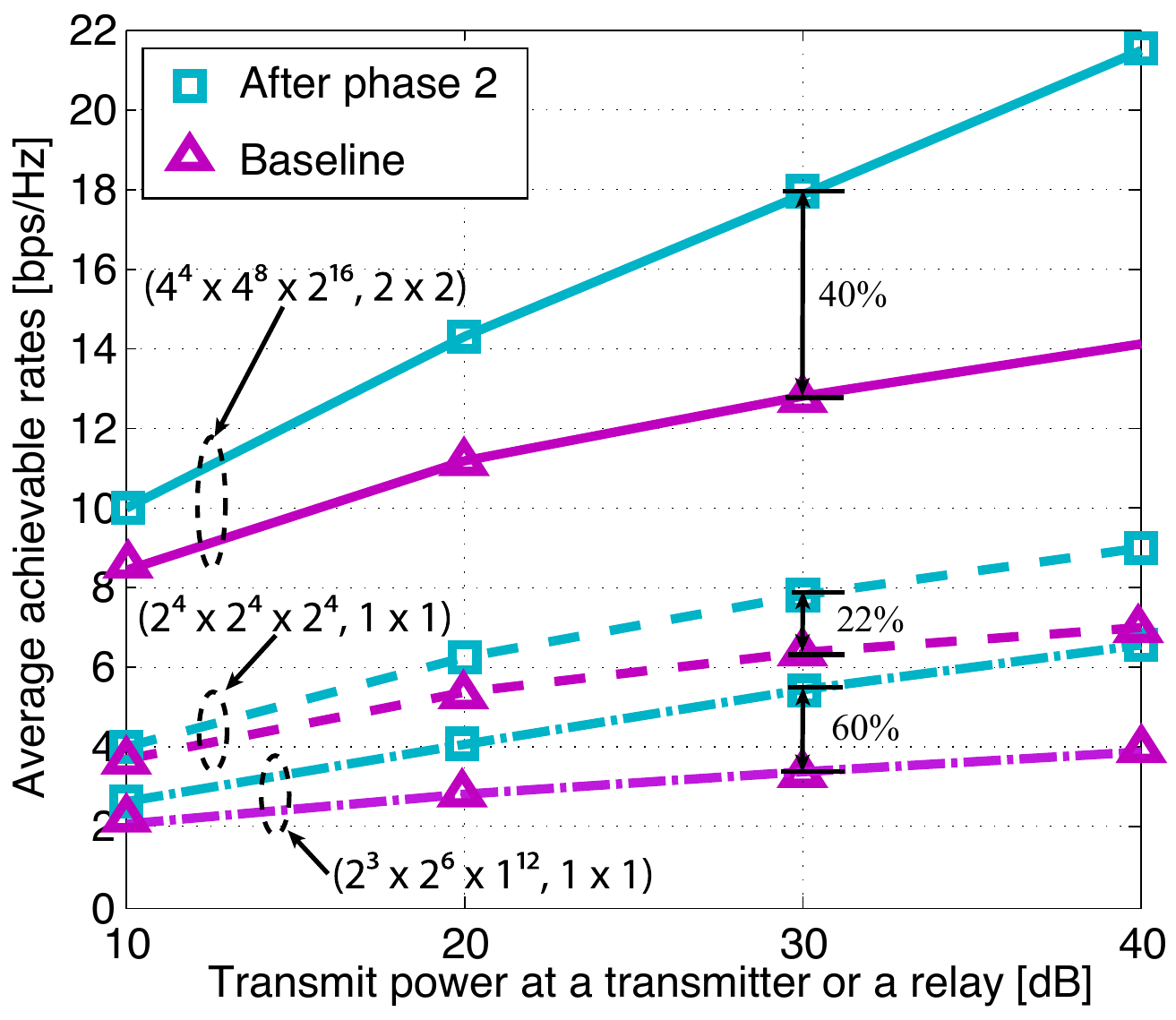}
\caption{Comparison of average end-to-end sum-rates of \emph{After phase 2} and \emph{Baseline} to show the benefits of the second phase of the proposed algorithm.}
\label{fig:ApproximateSRConf1and6and14}
\end{figure}

Note that the relative gains provided by each of the last two phases depend on the configurations. Fig. \ref{fig:ApproximateSRConf1and6and14} provides the simulation results for \emph{After phase 2} and \emph{Baseline} for the following three systems: $(2^{3}\times 2^{6}\times 1^{12}, 1 \times 1)$, $(2^{4} \times 2^{4} \times 2^{4}, 1 \times 1)$, and $(4^{4} \times 4^{8} \times 2^{16}, 2 \times 2)$. We observe that even before rate-matching power control, the proposed algorithm always outperforms the baseline. At $p_{\rT} = p_{\rX} = 30$ dB, the gain is 60$\%$ for the system $(2^{3}\times 2^{6}\times 1^{12}, 1 \times 1)$, 22$\%$ for the system $(2^{4} \times 2^{4} \times 2^{4}, 1 \times 1)$ and  40$\%$ for the system $(4^{4} \times 4^{8} \times 2^{16}, 2 \times 2)$.

\subsection{Opportunistic Solutions}
\begin{figure}[!h]
\centering
\includegraphics[width=3.4in]{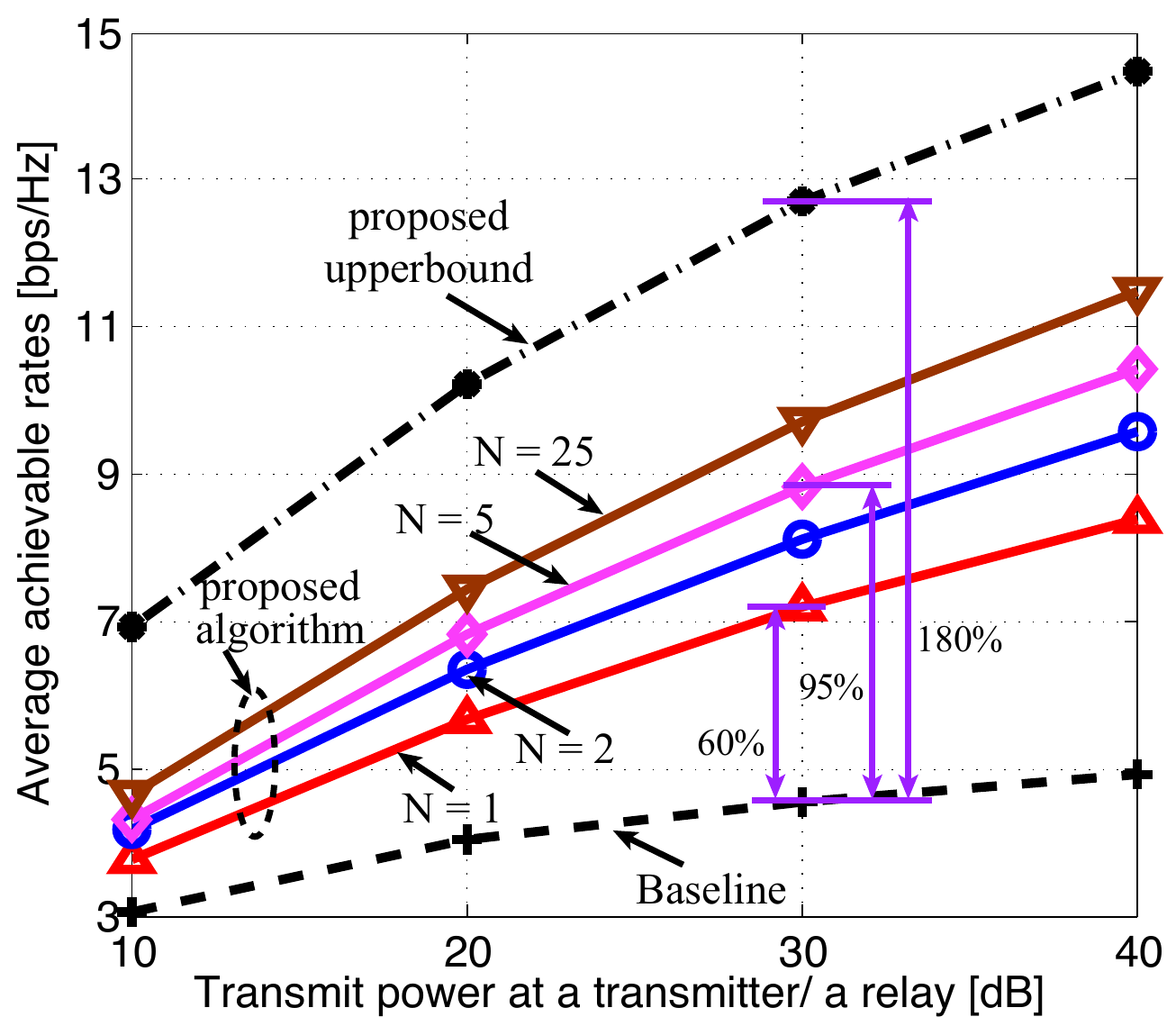}
\caption{Average end-to-end sum-rates of opportunistic solutions of the proposed three-phase algorithm for $N \in \{1, 2, 5, 25\}$ for $(2^{3} \times 2^{6} \times 2^{12}, 1 \times 1)$.}
\label{fig:opportunisticPCConf2}
\end{figure}

We adopt an opportunistic approach to improve the end-to-end sum-rate performance. Let $N$ be the number of random initializations in the opportunistic approach. The proposed algorithm is repeated $N$ times with the random initializations and choose the one with the highest achievable end-to-end sum-rates. Fig. \ref{fig:opportunisticPCConf2} provides the average achievable end-to-end sum-rates of the opportunistic solutions of the proposed three-phase algorithm with the number of initializations $N \in \{1, 2, 5, 25\}$ for the system $(2^{3} \times 2^{6} \times 2^{12}, 1 \times 1)$. To provide a benchmark, we also show the results for the fixed average normalized second-hop sum-rates, which provides an upper-bound for the solutions. As expected, increasing $N$ improves the end-to-end sum-rates achieved by the proposed algorithm with opportunistic implementation. At $p_{\rT} = p_{\rR} = 30$ dB, the opportunistic solution with $N = 5$ nearly doubles the end-to-end sum-rates when compared to the baseline. Nevertheless, the additional gains obtained by using an additional random initialization in the opportunistic solutions decreases in $N$. At $p_{\rT} = p_{\rR} = 30$ dB, the opportunistic solution with $N = 25$ achieves nearly $75\%$ the value of the upper-bound. 

\subsection{Varying Timesharing Values}
\begin{figure}[!h]
\centering
\includegraphics[width=3.4in]{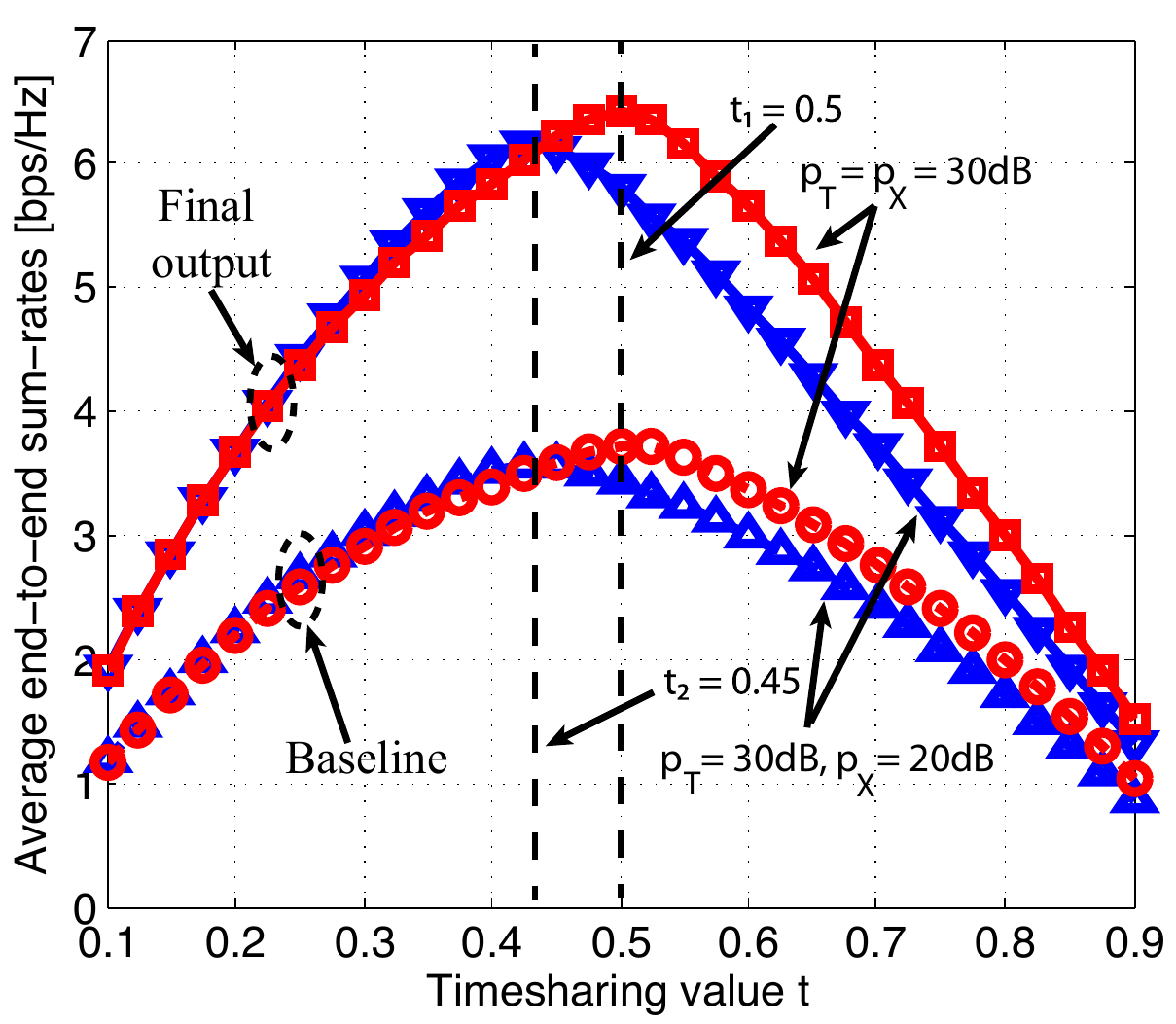}
\caption{Average end-to-end achievable rates as a function of timesharing $t$ for $(2^{4} \times 2^{8} \times 2^{16}, 1 \times 1)$.}
\label{fig:timesharing}
\end{figure}

Fig. \ref{fig:timesharing} presents the average achievable end-to-end sum-rates as functions of the timesharing values for the system $(2^{4} \times 2^{8} \times 2^{16}, 1 \times 1)$ for the following two cases: i) Case 1 with $p_{\rT} = p_{\rX} = 30$ dB and ii) Case 2 with $p_{\rT} = 30$ dB and $p_{\rX} = 20$ dB. We notice that for the proposed algorithm and the baseline in each case has the same optimal timesharing value, $t_{1} = 0.5$ for Case 1 and $t_{2} = 0.425$ for Case 2. These optimal timesharing values approximately equalize the average normalized sum-rates on two hops. This emphasizes the importance of matching the rates on two hops. In addition, we observe that thanks to the explicit consideration of $t$ for matching the rates on two hops, the proposed algorithm has large gains, between $50\%$ and $70\%$, over the baseline for $t \in [0.1, 0.9]$.

\section{Conclusions and Future Work}\label{sec:conclusion}
We proposed a cooperative algorithm for jointly designing transmit precoders at the transmitters and relays for the decode-and-forward relay interference broadcast channel to (approximately) maximize end-to-end sum-rates. The naive approach of applying single-hop interference management strategies separately for two hops of the system causes rate mismatch, leading to low end-to-end sum-rates. The main challenges were interference management and rate mismatch between the rates on the two hops. Our solution is a three-phase cooperative algorithm. The first phase focuses on the design of precoders at the relays to maximize the second-hop sum-rates. The second phases uses the knowledge of the timesharing value and the second-hop achievable rates to design the precoders at the transmitters to maximize an approximate of the end-to-end sum-rates. The last phase uses power control at the transmitters to eliminate any residual rate mismatch to further improve the end-to-end sum-rates. The proposed algorithm allows for distributed implementation and has fast convergence behavior, making it suitable for practical systems. Simulations showed that the proposed algorithm obtains much higher end-to-end sum-rates than the naive approach.

The work assumes instantaneous and perfect CSI is available at the transmitters and relays. Of course, perfect CSI is not realizable in practical systems where there is channel estimation error, time-variation in the channel, thermal noise, errors on the feedback link, and signal processing delay. Future work could investigate and quantify impacts of imperfect CSI, e.g., due to channel estimation or CSI feedback delay, on the proposed algorithm. Based on this knowledge, future work should focus on developing algorithms that are robust in presence of CSI uncertainty. Our proposed algorithm provides benchmarks for future work that accounts for more practical impairments.

\bibliographystyle{IEEEtran}
\bibliography{IEEEabrv,TVTrelayIBC}

\end{document}